\documentclass[12pt,preprint]{aastex}
\begin{document}

\parindent=1.0cm

\baselineskip=0.8cm

\title{GeMS in the Outer Galaxy: Near-Infrared Imaging of Three Young Clusters at Large Galactic Radii\altaffilmark{1}}

\author{T. J. Davidge}

\affil{Dominion Astrophysical Observatory,
\\National Research Council of Canada, 5071 West Saanich Road,
\\Victoria, BC Canada V9E 2E7}

\altaffiltext{1}{Based on observations obtained at the Gemini Observatory, which is
operated by the Association of Universities for Research in Astronomy, Inc., under a
cooperative agreement with the NSF on behalf of the Gemini partnership: the National
Science Foundation (United States), the National Research Council (Canada), CONICYT
(Chile), the Australian Research Council (Australia), Minist\'{e}rio da Ci\^{e}ncia,
Tecnologia e Inova\c{c}\~{a}o (Brazil) and Ministerio de Ciencia, Tecnolog\'{i}a e
Innovaci\'{o}n Productiva (Argentina).}

\begin{abstract}

	Images recorded with the Gemini South Adaptive Optics Imager (GSAOI) 
and corrected for atmospheric seeing by the Gemini Multiconjugate 
Adaptive Optics System (GeMS) are used to investigate the stellar contents of the young 
outer Galactic disk clusters Haffner 17, NGC 2401, and NGC 3105. 
Ages estimated from the faint end of the main sequence (MS) and the 
ridgeline of the pre-main sequence (PMS) on the $(K, J-K)$ color-magnitude diagrams 
are consistent with published values that are based on the MS turn-off, with the 
GSAOI data favoring the younger end of the age range for NGC 2401 in the literature. 
The mass function (MF) of NGC 2401 is similar to that in the Solar neighborhood, 
and stars spanning a wide range of masses in this cluster have 
similar clustering properties on the sky. It is concluded that 
NGC 2401 is not evolved dynamically. In contrast, the MF of 
Haffner 17 differs significantly from that in the Solar neighborhood 
over all masses covered by these data, while the 
MF of NGC 3105 is deficient in objects with sub-solar masses when compared with the 
Solar neighborhood. Low mass objects in Haffner 17 and NGC 3105 are also more uniformly 
distributed on the sky than brighter, more massive, MS stars. This is consistent 
with both clusters having experienced significant dynamical evolution.

\end{abstract}


\section{INTRODUCTION}

	It is generally believed that stars do not form in isolation, although the 
clusters and associations in which they form are subject to disruption by a combination 
of internal (e.g. winds and supernovae) and external (e.g. 
interactions with giant molecular clouds) processes. Evidence 
that most clusters are short-lived comes from studies of the disks of nearby 
galaxies, which reveal that the majority of stars belong to the so-called 
`field' (e.g. Silva-Villa \& Larsen 2011). While a few clusters in disk environments 
survive their first few Myr, the pace of cluster disruption continues at a rapid pace, 
with a destruction rate of roughly an order of magnitude per decade in 
age for systems older than 10 Myr (Fall \& Chandar 2012).

	Young star clusters at large Galactocentric radii (R$_{GC}$) are of interest 
as they probe evolution in an environment that is systematically different from the 
Solar neighborhood. The relatively low density of the outer Galactic disk 
will affect the pace of cluster evolution that is driven by external processes, with 
the result that internal processes may play a larger role in the structural evolution 
of clusters at large R$_{GC}$ than in the Solar neighborhood. In addition, 
the metallicity of young objects tends to drop with increasing R$_{GC}$, and 
a lower metallicity may allow protostars in outer disk clusters to 
accrete material for longer periods of time than in the solar neighborhood, 
thereby altering the stellar MF near what will eventually 
become the bottom of the MS (e.g. De Marchi et al. 2011). 
Finally, if galactic disks form from the inside-out then clusters 
that are dissolving near the periphery of the present day disk 
may be contributing a significant fraction of the local field star 
component -- studies of such clusters might then yield insights into 
the assembly of the field star population at large R$_{GC}$. 

	In the present paper, images corrected for atmospheric distortions by 
the Gemini South Muti-conjugate Adaptive Optics System (GeMS) and 
recorded with the Gemini South Adaptive Optics Imager (GSAOI) are used to probe the 
near-infrared photometric properties and spatial distributions of MS 
stars and PMS objects in three young clusters in the outer Galaxy: Haffner 17, 
NGC 2401, and NGC 3105. These clusters were selected for this program 
based on their (1) location outside of the Solar Circle, (2) age ($\leq 0.1$ Gyr), 
(3) distance ($\geq$ a few kpc), and (4) angular size (r$\leq 1.5$ arcmin).
Near-infrared observations of these clusters are 
of interest since they are at low Galactic latitude, and so they may be subject to 
significant amounts of extinction. The high angular resolution delivered by 
GeMS is useful for overcoming the crowding that might be an issue at faint 
magnitudes while also allowing wide binary systems to be resolved. The latter are of 
interest as the binary frequency in a cluster provides insights into its dynamical state 
and the population of isolated low mass objects that will ultimately find their 
way into the field (e.g. Marks \& Kroupa 2011). 

	The ages, distance moduli ($\mu_0$), and reddenings of the clusters, 
taken from the August 2013 version of the WEBDA (Mermilliod 1995) database, are 
listed in Table 1. Also listed are $K$ magnitudes and radii obtained from 2MASS images. 
The magnitudes and radii are total values estimated from growth curves, and the 
integrated magnitudes of all three clusters are subject to significant sources of 
uncertainty. NGC 2401 is a diffuse cluster, and its total brightness 
is susceptible to subtle variations in the sky background in the 2MASS images.
As for Haffner 17 and NGC 3105, the integrated brightnesses of both clusters
are affected by bright stars near the cluster centers. In the case of NGC 3105 
the total $K$ magnitude drops by 2.1 magnitudes if the light from the 
two brightest stars is excluded. 

	In the only published photometric study of Haffner 17, 
Pedredos (2000) used $UBV$ CCD measurements to find $E(B-V) = 1.26 \pm 0.04$, 
$\mu_0 = 12.3 \pm 0.2$, and an age of 50 Myr. Comparisons with isochrones lead 
Pedredos (2000) to suggest that Haffner 17 may have a super-solar metallicity, 
although this is based on two red stars that may or may not be cluster members. 
If this metallicity is verified spectroscopically then Haffner 17 would be 
a rare example of a young metal-rich cluster at large R$_{GC}$.

	NGC 2401 has been the subject of three recent studies that have yielded a 
range of ages. Sujatha et al. (2004) present a CMD that shows a 
well-defined MS turn-off (MSTO) and numerous candidate evolved stars. 
They find an age of 0.1 Gyr, a distance of 3.5 kpc, and 
$E(B-V) = 0$. In contrast, Baume et al. (2006) measure an age of 25 Myr, 
a distance of $6.3 \pm 0.5$ kpc, and E$(B-V) = 0.36$. 
Significantly, Baume et al. (2006) also detect 
a population of possible PMS objects with $V \geq 17 - 18$, and a Be star that may 
be a cluster member -- both of these findings are consistent with a younger age 
than found by Sujatha et al. (2004). Most recently, Hasegawa 
et al. (2008) find a distance of 5.6 kpc for NGC 2401, with an 
age of 0.1 Gyr and $E(V-I) = 0.3$, which corresponds to 
$E(B-V) \sim 0.24$. A curious aspect of the Hasegawa et al. (2008) study is that the 
isochrones in their Figure 2 do not match the CMD near the MSTO, in the sense 
that there are blue stars with MS-like colors that form a well-defined sequence above 
the MSTO of the 100 Myr isochrone. If these blue stars are actual 
members of NGC 2401 and define the MSTO, as opposed to blue stragglers or field stars 
with fortuitous photometric properties, then the age estimated 
for NGC 2401 by Hasegawa et al. (2008) is too old.

	NGC 3105 has one of the largest R$_{GC}$ of any young cluster yet discovered 
in the Galactic disk. Early studies by Moffat \& Pim Fitzgerald (1974) and 
Pim Fitzgerald et al. (1977) found evidence of numerous Be stars and 
supergiants, pointing to a young age. Sagar et al. (2001) measured $E(B-V) = 
1.06$, a distance of $9.5 \pm 1.5$ Mpc, and an age of $25 \pm 10$ Myr. 
More recently, Paunzen et al. (2005) found $E(B-V) = 0.95 
\pm 0.02$, a distance of $8.53 \pm 1.03$ kpc, and an age log(t)$=7.30 \pm 0.1$ 
(i.e. t$= 20^{+5}_{-4}$ Myr).

	The data discussed here are used to examine the $(K, J-K)$ 
CMDs of the clusters, with specific emphasis on the magnitude interval that 
contains PMS objects and the the lower reaches of the MS. 
The ages of young clusters can be estimated using the 
brightness of the MS cut-off (MSCO) and the position of the PMS sequence on the CMD. 
Assuming a standard MF then this part of the CMD is less susceptible to the 
stochastic effects that can affect the properties of the more massive cluster 
members that define the MSTO. Insights can also be gleaned 
into the dynamical state of the clusters and -- possibly -- their 
early evolution by examining the $K$ luminosity function (LF) and 
spatial distributions on the sky of their component stars. 

	The observations are summarized in 
Section 2, while the reduction of the data and the techniques used to 
make the photometric measurements are discussed in Section 3. 
The CMDs are presented and examined in Section 4, while the LFs and spatial distributions 
of stars within the clusters are the subjects of Sections 5 and 6. 
The paper closes in Section 7 with a discussion and summary of the results.

\section{OBSERVATIONS}

	The data were recorded in January and February 2013 as part of program 
GS-2012B-SV-409. Details of the observations are summarized in Table 2. This program 
utilized two new Gemini capabilities -- the GeMS adaptive optics system and the 
GSAOI -- and these are briefly described below.

	GeMS (Rigaut et al. 2012; Neichel et al. 2012) is the facility adaptive optics 
(AO) system on Gemini South. GeMS monitors wavefront distortions using information 
provided by five laser guide stars, up to three natural guide stars (NGSs), and an 
on-detector guide source. The result is a seeing-corrected signal that covers a much 
larger angular area on the sky than is delivered by -- say -- a 
single beacon AO system that is conjugated to the telescope primary mirror.

	The brightnesses and locations of the NGSs in the GeMS field 
play a key role in determining the degree of correction and the angular stability of the 
point spread function (PSF). In sparse stellar environments 
the choice of NGSs may be problematic, with selection based on 
the few available objects that are bright enough to serve as NGSs. Fortunately, 
the rich population of bright cluster members and field stars 
provided a number of potential NGSs for the present program. Three NGSs 
that defined a more-or-less equilatoral triangle-shaped asterism were identified for 
each cluster, and {\bf the mean $R$ magnitudes of these, $<R>$, are listed in Table 2.} 
While the degree of correction and the stability of the 
PSF across the field depends on a number of factors, such as guide star brightness and 
the atmospheric conditions in the dominant turbulent layers, an equilatoral 
triangle-shaped asterism should deliver a near-optimum level of PSF uniformity 
across the GSAOI field.

	The wavefront-corrected beam from GeMS was imaged by GSAOI (McGregor et al. 
2004; Carrasco et al. 2012), which was designed specifically for 
use with GeMS. The detector in GSAOI is a mosaic of four $2048 \times 2048$ 
HgCdTe arrays that cover an $85 \times 85$ arcsec area with 0.02 arcsec 
pixel$^{-1}$ sampling. Gaps between the arrays
form a cross-shaped pattern with 3 arcsec wide arms on the sky. 

	Images were recorded through $J$, $Ks$, and Br$\gamma$ filters, with 
short and long exposure times to allow a wider 
magnitude range to be sampled than is possible with only a single exposure time.
A three point (short exposure) and five point (long exposure) 
linear north/south dither pattern was employed that allowed one arm of the gap on the 
detector mosaic to be filled. {\bf The cluster fields are not 
crowded, and so the seeing requirements for this program were set to 85\%ile. The 
Fried (1966) parameter, $r_0$, provides one measure of image quality. The 
$r_0$ measurement taken from the image header of the exposure that was recorded midway 
through the observing sequence for each cluster is listed in Table 2. The seeing degrades 
as $r_0$ moves to smaller values. During a seeing campaign in 1998, Vernin et al. 
(2000) found a mean $r_0$ of 19 cm on Cerro Pachon, with a minimum $r_0$ of 5 cm, and a 
maximum of 40 cm. The values of $r_0$ listed in Table 2 are less than the mean found 
by Vernin et al. (2000), indicating that these measurements were made during 
worse than normal seeing conditions, as was intended.}

	The mean FWHM values, $<FWHM>$, and the standard deviation about these means, 
$\sigma_{FWHM}$, are also listed in Table 2. These gauge the level of correction 
and the uniformity of the PSF across the GSAOI field, and the entries in Table 2 were 
measured from the PSF stars that were used in the photometric analysis of the long 
exposure images (Section 3). The PSF stars in the Br$\gamma$ images have similar FWHMs 
to those in $Ks$, and so $<FWHM>$ and $\sigma_{FWHM}$ are not listed for that filter. 

	The angular resolution delivered by an AO system during given seeing 
conditions depends on wavelength, in the sense that as one moves to longer wavelengths 
then the corrected angular resolution will approach the telescope diffraction limit 
at that wavelength. At the same time, the PSF will also become more uniform across 
the field (i.e. $\sigma_{FWHM}$ in Table 2 should decrease). 
The entries in Table 2 show that both the angular resolution and stability of the FWHM 
across the field is better in $Ks$ than in $J$, as expected.

	{\bf The telescope diffraction limit, $1.2 \times \frac{\lambda}{D}$, 
where $D$ is the diameter of the primary mirror, is 0.04 arcsec in $J$, and 
0.07 arcsec in $Ks$. The $Ks$ images of NGC 2401 have angular resolutions that come 
closest to the system diffraction limit. The entries in Table 2 indicate 
that GeMS can deliver $\sim 0.1$ arcsec angular resolution images in $J$ and $Ks$ 
with a stability in the FWHM of 10\% across the GSAOI field even when the imaging 
conditions are less than ideal.}

\section{DATA REDUCTION \& PHOTOMETRY}

\subsection{Producing Final Images}

	The initial processing of the data followed standard procedures for 
near-infrared imaging. The basic steps were (1) dark subtracton, (2) flat-fielding, 
and (3) the removal of thermal emission artifacts that are produced by warm objects 
along the optical path, such as dust on the GSAOI entrance window. The first two steps 
used calibration frames that were recorded specifically for these 
purposes. As for the third step, calibration frames were constructed 
by median-combining the flat-fielded long exposure images of all program clusters 
on a filter-by-filter basis after these were processed to correct for differences 
in exposure time and sky level (see below). The resulting calibration 
frames are not `sky flats', as they are constructed from images that have already been 
corrected for flat-field variations. Rather, they monitor an 
additive component in the background that constitutes a significant source of noise.

	To construct the background calibration frames it is necessary to 
create a homogeneous set of images that have a common effective exposure time and 
have been corrected for variations in the background sky level, 
which can be significant over timescales of a few minutes in the near-infrared. 
The individual cluster images were first normalized to a common exposure time 
(1 second in this case). A constant sky level, measured by taking the mode 
of pixel intensities in each frame, was then subtracted from each image. 
As four different clusters were observed and the data were recorded with on-sky 
dithering, then the median intensity of all images at each
GSAOI detector pixel is expected to be free of contamination from 
bright and moderately bright sources, leaving only thermal emission signatures. 
This expectation was borne out by visual inspection of the background calibration frames.

	The penultimate processing step is to correct for distortions 
that are introduced by the GeMS and GSAOI optics, and are manifest as a change 
in pixel scale with location in the images. These distortions are primarily (but
not exclusively) radial in nature and are most obvious near the edge of the science 
field in images that are registered near the center of the GSAOI field. They can appear 
either as filter-dependent diffences in source centroids or as differences between 
source centroids in images recorded in the same filter but at different points in 
the on-sky dither pattern. 

	When comparing observations recorded in different filters, the offsets 
are largest when $J$ and $Ks$ images are compared, where they can amount 
to differences in image centroids of a tenth of an arcsec (i.e. a few GSAOI pixels) 
near the edge of the GSAOI science field. Offsets of 
a few hundredths of an arcsec (i.e. a GSAOI pixel) are typical when 
Br$\gamma$ and $Ks$ images are compared. To obtain a common reference grid 
for image combination and photometric analysis, the $J$, Br$\gamma$, and dithered 
$Ks$ images were geometrically transformed using the IRAF Geomap and Geotran tasks
into the reference frame defined by the $Ks$ image recorded 
at the null position of the dither pattern. The final images for each 
cluster were then found by taking the median signal level on a filter-by-filter basis 
at each pixel location in the transformed and aligned images. 

\subsection{Photometric Measurements, Calibration, and Characterization}

	Stellar brightnesses were measured with the PSF-fitting program ALLSTAR 
(Harris \& Stetson 1988). PSFs for each cluster were constructed from between 
15 to 20 stars using the $PSF$ routine in the DAOPHOT (Stetson 1987) package. 
The same stars were used to construct the PSFs in all 
three filters. The wings of AO-corrected PSFs typically extend out to 
radii that are comparable to the uncorrected seeing disk, and a 1 arcsec 
PSF extraction radius was used here.

	PSF stars were selected based on brightness and the absence of obvious 
companions. An additional criterion was location in the GSAOI science field, with the 
intent of obtaining a more-or-less uniform distribution of PSF stars across the 
$85 \times 85$ arcsec$^2$ field. Monitoring the PSF at various points across the science 
field is important as modest residual PSF variations remain in the GeMS-corrected beam. 
The $\sigma_{FWHM}$ entries in Table 2 indicate that there are variations in the 
FWHM of PSF stars of up to a few hundredths of an arcsec, and a visual inspection of 
the FWHM values over the GSAOI field indicated that 
modest systematic trends were evident. In fact, 
photometry obtained with a PSF that has a first order spatial variation delivered 
tighter sequences on the CMDs than those that resulted from the application of a 
constant PSF. Still, the differences between the CMDs that result 
from the application of fixed and variable PSFs are not large, and the basic results 
of this study would not change substantially if fixed PSFs were used to 
obtain the photometry. 

	{\bf The $J$ and $Ks$ photometry was calibrated using 28 observations 
of 13 Persson et al. (1998) standard stars that were recorded throughout the GeMS SV run. 
The $1\sigma$ formal error in the $J$ and $K$ zeropoints is $\pm 0.04$ magnitude, with 
a standard deviation about the mean of $\pm 0.13$ magnitudes. 
The calibration was checked against photometric measurements 
made from 2MASS images, although the vastly different angular resolutions 
and faint limits of the GSAOI and 2MASS observations limits the number of 
suitable stars that can be used in such a comparison. The GSAOI and 2MASS images were 
also recorded at different epochs, and so stellar variability is a potential source 
of scatter. The brightnesses of sources in the 2MASS images of the clusters 
were measured with ALLSTAR, and the results were calibrated using relevant information 
in the 2MASS image headers. Restricting the comparisons to isolated 
objects with $K < 14.5$, then the mean differences between 
the two datasets are $\Delta(K) = 0.06 \pm 0.05$ and $\Delta(J-K) = -0.08 
\pm 0.09$.  These differences are in the sense GSAOI--2MASS, and the quoted 
uncertainties are $1\sigma$ errors in the mean. It is evident that the GSAOI and 
2MASS measurements are in a consistent photometric system. A similar conclusion was 
reached by Davidge et al. (2013), based on comparisons of GSAOI and 2MASS $(K, J-K)$ 
CMDs of the cluster Haffner 16, which was also observed as part of program 
GS-2012B-SV-409.} 

	The calibration of the Br$\gamma$ images is 
tied to the $Ks$ calibration. The central wavelengths and 
the mean throughputs of the Br$\gamma$ and $Ks$ filters are not vastly different, 
and a zeropoint for the $Br\gamma$ filter was found by scaling the $Ks$ 
zeropoint to compensate for the difference in the effective bandpasses between the 
two filters. As shown in the next section, this scaling results in Br$\gamma - 
Ks \sim 0$ for most sources.

	{\bf Artificial star experiments were run to determine sample 
completeness and to characterize uncertainties in the 
photometry. The artificial stars were constructed by scaling the PSFs that were used 
to make the photometric measurements and were assigned $J-K$ and $Br\gamma-K$ colors 
that are representative of the cluster sequences on the CMDs. As with real stars, 
an artificial star was considered to be recovered only if it was detected in two 
filters (either $J + K$ or $Br\gamma + K$). The artificial star experiments indicate that 
the completeness fractions near the faint end of the CMDs vary from cluster-to-cluster, 
as expected due to differences in angular resolution. The 
$(K, J-K)$ CMDs of all three clusters are complete when $K \leq 19.5$.}

\section{COLOR-MAGNITUDE DIAGRAMS}

\subsection{A First Look at the $(K, J-K)$ CMDs}

	The $(K, J-K)$ CMDs of the clusters are presented in Figures 1 -- 
3. {\bf The blue error bars in the left hand panels show the $\pm 1\sigma$ random 
uncertainties in $J-K$ computed from the artificial star experiments. The 
predicted dispersion more-or-less matches the observations at $K=19$ and 
$K=20$, suggesting that the scatter af the faint end of the CMDs is 
dominated by photometric errors. In contrast, at $K \leq 18$ the random uncertainties 
tend to be narrower than the observed scatter, suggesting that the width of the 
CMDs at these magnitudes reflects an intrinsic dispersion in the photometric 
properties of objects.} 

	The CMD of Haffner 17 has a moderately tight 
sequence with $K < 18$ that is dominated by cluster stars. The scatter that 
is present at most magnitudes is likely due to a mix of cluster members, which 
are a combination of main sequence and PMS objects, and field stars. Star counts 
from the Robin et al. (2003) model suggest that the fraction of field stars 
in the Haffner 17 CMD ranges from 30\% at the bright end to 70\% at the faint end.
These numbers were computed using the full range of stellar types available 
for the Robin et al. (2003) model, with the reddening assumed to be
uniformly distributed along the line of sight towards the cluster. 
These assumptions are used for all star count models discussed throughout this work.

	The dominant sequence of objects on the NGC 2401 CMD is not as well-defined 
as in the other two clusters. Baume et al. (2006) present a near-infrared 
CMD of NGC 2401 obtained from 2MASS images that also shows substantial scatter. 
The dispersion along the color axis is likely not due to differential reddening. 
If this were the case then $\Delta E(B-V) \sim 0.4 - 0.6$ 
magnitudes across the GSAOI field. Differential reddening of this size would be 
surprising given that stars in NGC 2401 have an average $E(B-V) = 0.15$ (Section 4.2). 
There are also no obvious dust lanes in the DSS or 2MASS images of NGC 2401. 
The star count models of Robin et al. (2003) suggest that 25 -- 30\% of the 
objects in the CMD of NGC 2401 are field stars, and we suspect that contamination 
from these objects is the dominant source of the scatter along the $J-K$ axis.

	NGC 3501 has the most richly populated CMD in our sample. 
A locus of cluster stars can be seen, although there is 
substantial scatter. Comparisons with star count models from Robin et al. 
(2003) suggest that field stars account for 40\% of objects near 
the bright end of the NGC 3105 CMD and 60\% of the objects near the faint end. 
Hence, much of the scatter is due to field stars. Of particular interest for 
the current work is the sequence of objects with $J-K \sim 1$ that runs from $K = 
17.5$ to $K = 19$. Comparisons with isochrones indicate that this 
sequence is populated by PMS objects that belong to NGC 3105 (Section 4.2). 

\subsection{Comparisons with Isochrones}

	The CMDs are compared with Padova Isochrones 
in the middle and right hand panels of Figures 1 
-- 3. These isochrones, which are described by Bressan et al. (2012),
are of interest for the present work as they include PMS evolution. The isochrones 
used here have a metallicity $Z = 0.016$, which provides a consistent point 
of comparison with previous studies. The consequences of adopting different metallicities 
on the distance, reddening, and age estimates are examined in the next section.

	The comparisons in the middle panels of Figures 1 -- 3 use reddenings 
and distance moduli from Pedredos (2000) for Haffner 17, Baume et al. (2006) 
for NGC 2401, and Paunzen et al. (2005) for NGC 3501. $E(J-K)$ was computed from 
the published $E(B-V)$ values using the relations in Table 6 of Schlegel et al. (1998).
It can be seen from the middle panels of Figures 1 -- 3 that 
the published reddenings and distance moduli can be adjusted to achieve better 
agreement between the isochrones and the observations. To this end, independent 
reddenings, distances, and ages were found by comparing the properties of 
MS and PMS stars in the clusters with the isochrones. These revised parameters, 
which were used to make the comparisons in the right hand panels of Figures 
1 -- 3, are discussed in the following sections.

\subsubsection{Reddenings and Distance Moduli}

	$E(J-K)$ was found by matching the models to the blue edge of the 
near-vertical upper MS segment on the CMDs. There is also 
a vertical plume of faint red objects that is most clearly defined 
in the NGC 2401 CMD in Figure 2, which is another potential probe of reddening. 
However, the isochrones indicate that this 
feature is populated by a mix of field and cluster stars, and so 
gives only a loose check on cluster reddening. 

	In order to make comparisons with published reddenings, $E(B-V)$ was calculated 
from $E(J-K)$ using the reddening relations listed in Table 6 of Schlegel et al. (1998). 
Random uncertainties were estimated by perturbing the adopted reddening and determining 
the smallest change in this quantity that noticeably degraded 
the agreement between the isochrones and the observations. These 
experiments suggest that the random uncertainties are $\pm 0.02$ magnitudes 
in $E(J-K)$, and $\pm 0.03$ magnitudes in $E(B-V)$. Departures from the reddening law 
used here are an obvious source of systematic uncertainty in the $E(B-V)$ values. 

	The CMDs of systems with ages older than $\sim 20$ Myr have a pronounced bend 
on the MS near M$_K = 2$. This feature, which is due to 
a change in opacity in stellar atmospheres, is employed here to determine distance. 
As with the reddenings, random uncertainties in the distance modulus were found by 
perturbing this quantity and assessing the impact on the agreement between the 
isochrones and the observations. These experiments suggest that the random uncertainty 
in the observed distance modulus is $\pm 0.05$ magnitudes. 
With the reddening and distance set then an age can be estimated using the 
point on the CMD where the dominant stellar sequence departs from the MS 
as defined by old stellar systems and -- for NGC 2401 
and NGC 3105 -- from a direct comparison between the PMS sequence and isochrones. 
Ages are estimated in Section 4.2.2.

	The reddenings and distance moduli obtained from the GSAOI images 
do not show systematic effects. The reddening estimated by 
Predredos (2000) for Haffner 17 must be reduced to match the 
$(K, J-K)$ CMD, while $\mu_0$ must be increased. In contrast, 
for NGC 2401 a higher $E(B-V)$ and a lower $\mu_0$ are required to get good 
agreement with the models. Finally, for NGC 3105 the near-infrared CMD calls for 
both a lower $E(B-V)$ and a lower $\mu_0$.

	There is a metallicity gradient among young objects in the Galactic disk, 
and the clusters discussed here may have a sub-solar metallicity given their R$_{GC}$. 
Balser et al. (2011) find an azimuthally-averaged radial 
O/H gradient of --0.045 dex kpc$^{-1}$ in the Galactic disk, and NGC 3105 would 
then have [O/H] $= -0.3$ dex if its metallicity follows this general trend. 
Still, in the absence of spectroscopic data, efforts to estimate 
metallicity based only on location in the Galaxy are frustrated by the variation 
in metallicity at a fixed R$_{GC}$, which is compounded by variations tied to the 
Galactic position angle (e.g. Balser et al. 2011). If the range in 
gradients found in the Balser et al. (2011) study are adopted as representative of the 
entire disk then [O/H] for NGC 3105 would fall between --0.2 and --0.5 dex. 

	To assess the result of adopting different metallicites on the cluster 
parameters estimated from the GSAOI data, the $(K, J-K)$ 
CMD of NGC 3105 is compared with non-solar metallicity 
isochrones in Figure 4. NGC 3105 was selected for this comparison because it 
has the largest R$_{GC}$, making it the cluster most 
likely to have a non-solar metallicity. Comparisons are made with $Z = 0.010$ and 0.025 
models, which is the range in metallicity that more-or-less reflects the $\pm 0.2$ dex 
scatter that is seen in the metallicity $vs$ R$_{GC}$ curve defined by Galactic 
HII regions (e.g. Balser et al. 2011).

	The comparisons in Figure 4 indicate that $E(B-V)$ is not sensitive to 
isochrone metallicity. This is not surprising given that the reddening is 
determined from bright MS stars with photospheres in which energy transport 
is via radiative processes and electron scattering is the primary source of opacity. 
On the other hand, the Z=0.010 and Z=0.025 isochrones give $\mu_0$ values that differ 
by 0.3 dex, which corresponds to 1 kpc at the approximate distance of NGC 3105. This 
reflects the metallicity sensitivity of stellar structure near the M$_K = 2$ 
CMD feature. 

\subsubsection{Ages}

	The luminosity of the MSTO is the classical benchmark for age estimates. 
However, the upper part of the MS may be poorly populated in young, 
low mass clusters. Statistical flucuations amongst the most massive 
stars in low mass clusters may even introduce a systematic bias in 
age estimates, in the sense that ages may be overestimated on average. 
In contrast, the MSCO and PMS sequences are in parts of the CMD that -- 
barring an anomalous MF -- are more richly populated than the upper portions of the MS. 
The MSCO and PMS thus provide a means of gauging the ages of young and intermediate-age 
low mass clusters that is less susceptible to statistical flucuations in stellar 
content than the MSTO.

	Figure 5 shows the portions of the $(K, J-K)$ CMDs that cover the faintest MS 
stars and brightest PMS objects. The yellow line is a 1 Gyr isochrone that is included 
to show a reference `old' MS. Ages are estimated using two criteria: (1) the magnitude 
at which the majority of stars at the blue edge of the data envelope depart from the 1 
Gyr isochrone (i.e. the MSCO), and (2) direct comparisons between the isochrones and 
PMS sequences. It should be recalled that the age estimates made from PMS sequences 
are not sensitive to $\pm 0.2$ dex changes in metallicity (Figure 4).

	The modest number of sources in Haffner 17, coupled with the sizeable 
contamination from field stars, complicates efforts to estimate its age. 
The upper and lower limits of the MSCO in this cluster are indicated 
in the left hand panel of Figure 5. The faint limit of the MSCO is the point 
below which sources no longer fall within $\pm 0.1$ magnitude of the 1 Gyr MS 
along the $J-K$ axis, and is defined by the 5 points between $K = 17.5$ 
and 17.65 that lie close to the 1 Gyr MS. {\bf This $\pm 0.1$ magnitude criterion 
in $J-K$ exceeds the estimated random uncertainties in the photometry, and so 
provides a conservative means of assessing when stars are not on the MS.} 
Looking at brighter magnitudes, there is a $\sim 0.2$ magnitude gap between 
this clump of points and the next group of points that fall close to the MS, and these 
define the bright limit of the MSCO estimate. Objects that scatter closely about 
the MS are relatively common at brighter magnitudes.

	A well-defined PMS sequence is not evident in the CMD of Haffner 17, and this is 
likely a consequence of the flat LF of this cluster (Section 5) combined with 
the high levels of field star contamination. Therefore, the age estimate 
for this cluster is restricted to the MSCO, which suggests that Haffner 17 has an 
age in the range 45 -- 60 Myr. This age is consistent with that estimated by 
Prededos (2000) from the MSTO.

	Given the comparatively low level of field star contamination towards NGC 2401, 
the majority of points in the middle panel of Figure 5 likely belong to that cluster. 
The faint limit of the MSCO for NGC 2401 is based on the clump of four points 
near $K = 17$ that fall immediately to the right of the 1 Gyr MS. 
The next faintest clump of points that are close to the 1 Gyr MS are seen near 
$K = 17.9$, and this large gap indicates that $K = 17$ is a reasonable estimate 
of the faint limit of the MSCO, {\bf even after considering the random dispersion in 
$J-K$ estimated from the artificial star experiments}. Between $K = 17$ and 
17.2 there are two points that fall within a few hundredths of a magnitude 
in $J-K$ of the MS. The faintest of these at $K = 17.2$ 
defines the bright limit of the MSCO for this cluster.

	The age of NGC 2401 can also be estimated from PMS stars that populate 
the sequence of points between $K = 17$ and 18 and scatter about 
the 20 Myr isochrone. Given the modest level of field star contamination 
then the majority of these objects are likely members of NGC 2401. 
Considering the MSCO and PMS ages of NGC 2401 together, then the cluster has an age 
in the range 20 -- 30 Myr. This is consistent with the age found by Baume et al. (2006), 
but is much younger than the ages found by Sujatha et al. (2004) and Hasegawa et al. 
(2008). Assuming a normal MF -- an assumption that is justified in 
Section 5 -- then the older ages predicted by the latter two studies 
can not explain the absence of MS stars between $K = 17$ and $K = 17.9$ and the 
presence of objects as bright as $K = 17$ with PMS-like photometric properties. 

	Field star contamination is an issue for determining the MSCO magnitude 
in NGC 3105. Still, there are large numbers of points that fall along -- or 
immediately to the right of -- the 1 Gyr MS isochrone to as faint as $K = 17.7$. 
Data points clustering about the MS are present 
at fainter magnitudes, but they form a more diffuse sequence than 
at brighter magnitudes {\bf with a dispersion that exceeds that expected from 
random photometric errors alone}. This suggests that these objects are field stars. 
Therefore, $K = 17.7$ is adopted as the MSCO faint limit in NGC 3105. 
Keeping with the spirit of the procedure applied for Haffner 17 and NGC 2401, then the 
bright limit of the MSCO in NGC 3105 is assigned to the next brightest 
clump of points, which occur at $K = 17.5$. The age of NGC 3105 determined from the 
MSCO is then 25 Myr.

	An age for NGC 3105 can also be estimated from 
PMS stars that form a locus of points in the CMD with $K \geq 17.5$. 
The majority of points on this sequence fall between the 20 and 40 Myr 
isochrones and are indicative of an age near $\sim 25$ Myr, which is consistent 
with the age deduced from the MSCO. The age of NGC 3105 based on the MSCO and PMS 
is then $\sim 25$ Myr. This agrees with the ages found by Sagar 
et al. (2001) and Paunzen et al. (2006).

\subsection{Br$\gamma$ Emission}

	Young PMS stars that are accreting material may show Hydrogen 
emission lines in their spectra. The strength of this 
emission depends on the rate of accretion, which can vary with time. The overall 
accretion rate also depends on metallicity and age, in the sense that 
the mean accretion rate drops towards higher metallicities and older ages. 
Line emission is seen among PMS stars in the SMC with ages $\sim 20 - 30$ Myr 
(e.g. Figure 9 of De Marchi et al. 2011). Given that (1) the 
clusters discussed here are in the outer regions of the Galaxy, where there is 
a reasonable chance that they might have a sub-solar metallicity, and (2) there is 
evidence that NGC 2401 and NGC 3105 have ages of 20 -- 30 Myr, then it was decided 
to search for signatures of Br$\gamma$ emission from bright PMS stars.

	Br$\gamma$ emission lines in the spectra of 
very young PMS objects in the solar neighborhood have equivalent widths 
$\leq 10\AA$ (e.g. Donehew \& Brittain 2011). To detect 
similar emission lines in individual objects in the narrow-band Br$\gamma$ 
filter would require photometry that is reliable to within a few hundredths of a 
magnitude, and this is at the limits of the photometric uncertainties 
of even the brightest stars in the GSAOI images. 
However, the mean Br$\gamma - K$ colors of an ensemble of objects may be skewed 
systematically if Br$\gamma$ emission is present in a large number of its members. 
Davidge et al. (2013) investigate the mean Br$\gamma$--K colors of 
objects in the 10 Myr cluster Haffner 16, and find evidence of weak systematic 
Br$\gamma$ emission among PMS objects in that cluster.

	The ($K$, Br$\gamma - K$) CMDs of the clusters are shown in 
Figure 6, where the $\pm 1\sigma$ random errors in Br$\gamma - K$, calculated 
from the artificial star experiments, and the magnitude of the MSCO are indicated. 
The predicted random errors in $Br\gamma - K$ match the observed scatter near the faint 
ends of the CMDs. However, at intermediate magnitudes the observed scatter exceeds 
the predicted random errors, indicating that star-to-star differences in 
$Br\gamma - K$ are present.

	Sources with Br$\gamma$ in emission lie to the left of non-emitting 
sources on the $(K, Br\gamma - K)$ CMDs. The Br$\gamma - K$ color of Haffner 17 stays 
roughly constant with $K$ magnitude, with no obvious change in Br$\gamma - K$ 
near the MSCO. This suggests that Br$\gamma$ emission with equivalent widths 
of a few \AA\ is not present among the majority of PMS stars in Haffner 17.

	The situation is different for NGC 2401. In the magnitude 
interval $K = 14.5 - 16.3$ the majority of objects fall to the right of 
the Br$\gamma - K = 0$ line, whereas between $K = 16.3$ and 18 the majority of points 
fall to the left of this line. The artificial star experiments indicate that the data 
are complete in this magnitude range, and that the break is not due to a 
systematic error in the photometry. A Kolmogorov-Smirnov test indicates that the 
Br$\gamma-K$ distributions in the interval $K = 15.0 - 16.0$ differs from 
that in the interval $K = 16.5 - 17.5$ at roughly the 95\% confidence level. 
Thus, the Br$\gamma-K$ distribution changes within $\sim 0.6$ magnitude 
of the MSCO in NGC 2401.

	The morphology of the $(K, Br\gamma - K)$ CMD of NGC 3105 is similar 
to that of NGC 2401, but shifted $\sim 0.6 - 0.7$ magnitudes fainter in $K$. A 
jog can be seen near $K = 17$ in the NGC 3105 $(K, Br\gamma - K)$ 
CMD, in the sense that at magnitudes between $K = 16$ 
and 17 the majority of points fall to the right of the Br$\gamma - K = 0$ line, whereas 
between $K = 17$ and 18 the majority of points fall to the left of this line. 
This break occurs at a magnitude where the data are 100\% 
complete. A Kolmogorov-Smirnov test indicates that the Br$\gamma - K$ 
distributions of sources in the intervals $K = 16 - 17$ and $K = 17 - 18$ differ 
at roughly the 97\% confidence level. Therefore, as in NGC 2401, a change in the 
Br$\gamma - K$ colors of sources occurs $\sim 0.6$ magnitudes brighter than the MSCO.

	The breaks in the $(K, Br\gamma-K)$ CMDs of NGC 2401 and NGC 3105 occur near 
M$_K \sim 2.5 - 2.6$. Given that these clusters have similar ages, then that 
their $(K, Br\gamma - K$) CMDs show a feature at roughly the same 
M$_K$ suggests that a common mechanism may be responsible for this change. 
This being said, it is not clear if the change in the 
Br$\gamma - K$ distribution is due to the onset of Br$\gamma$ 
emission, or to a decrease in the strength of Br$\gamma$ absorption. 
PMS objects have lower surface gravities than MS stars with the 
same color, and so may have weaker Br$\gamma$ absorption line strengths.
The differences in the Br$\gamma$ properties of PMS and non-PMS sequence stars 
in NGC 2401 and NGC 3105 could be due to such a surface gravity effect. If this is the 
case then the Br$\gamma$ observations have picked up on a signature of PMS stars, 
albeit one that is due to a mechanism other than Hydrogen line emission.

	Could the break in the $(K, Br\gamma - K)$ CMDs not be related 
to PMS stars? The M$_K$ at which the change in $Br\gamma - K$ occurs corresponds to that 
of MS stars with spectral type mid-F, and the equivalent widths of hydrogen absorption 
lines change significantly between spectra types F0V (M$_K \sim 2$) to G0V (M$_K 
\sim 3$). However, a change in Br$\gamma - K$ colors due to spectroscopic variations
among MS stars would be expected to occur over $\sim 1$ magnitude in $K$, 
whereas the breaks found here seem to occur over a narrow magnitude range. 

	There is additional evidence that the changes in $Br\gamma - K$ are due to 
PMS stars. There is a well-populated, narrow PMS sequence on the $(K, J-K)$ CMD of 
NGC 3105, allowing PMS stars to be identified using their 
location on the CMD. The Br$\gamma - K$ colors of objects 
that fall along the PMS sequence can then be compared with those of objects 
that are not on this sequence, and so have a lower likelihood of being PMS stars. 
For the purposes of making such a comparison, objects 
in the interval $K = 17.5$ and 18.5 that fall within 
$\pm 0.15$ magnitude in $J-K$ of the PMS sequence 
on the $(K, J-K)$ CMD are assumed to be PMS objects, 
while those within the same $K$ magnitude range but having $J-K$ colors 
outside of the $\pm 0.15$ magnitude PMS strip are classified as non-PMS objects.

	If the PMS objects have Br$\gamma$ photometric properties 
that differ on average from those of non-PMS stars then there 
should be statistically significant differences in $<Br\gamma-K>$.
The 122 objects in the PMS sample have $<Br\gamma-K> = -0.002 \pm 0.006$, with 
a standard deviation $\sigma = \pm 0.060$. For comparison, the 66 
non-PMS objects have $<Br\gamma - K> = 0.057 \pm 0.026$ with 
$\sigma = \pm 0.221$. The uncertainties in $<Br\gamma-K>$ are the $1\sigma$ 
errors in the mean. The $<Br\gamma-K>$ of the PMS and non-PMS groups thus 
differ at the $2\sigma$ level, in the sense that the PMS objects have a 
smaller $<Br\gamma - K>$, and so are brighter in the Br$\gamma$ filter than objects 
in the non-PMS sample that have the same $K$. A $t$ test indicates that 
the means of the PMS and non-PMS samples differ at roughly 
the 98\% confidence level, whereas an $F$ test indicates that the standard 
deviations of these samples differ at more than the 95\% confidence level. 
These comparisons thus indicate that the objects on the PMS sequence in the 
$(K, J-K)$ CMD of NGC 3105 have photometric properties in the 
Br$\gamma$ filter that differ systematically from those of non-PMS objects in the 
same $K$ magnitude interval. The higher mean brightness in the Br$\gamma$ filter for 
PMS objects is consistent with either line emission or weaker Br$\gamma$ absorption. 

	The narrow-band Br$\gamma$ observations identify NGC 2401 and 
NGC 3105 as interesting targets for follow-up observations. 
If Br$\gamma$ emission is present then spectroscopy at visible/red wavelengths
should reveal wide-spread H$\alpha$ and H$\beta$ emission from PMS objects 
in these clusters. The detection of line emission in a large fraction of the PMS 
objects would be astrophysically interesting, as it would indicate that they are 
actively accreting material. Given that this activity would presumably 
have continued over the age of the clusters (20 -- 30 Myr), then the lowest mass 
MS stars that ultimately form in these clusters may be more massive than those that 
form in clusters where accretion is truncated after a shorter period of time. If 
NGC 2401 and NGC 3105 are typical young outer disk clusters 
then the low mass portion of the IMF in this environment 
may differ from that in the solar neighborhood.

\section{LUMINOSITY FUNCTIONS}

	The LFs of the clusters, corrected for 
field star contamination using number counts from the Robin et al. (2003) models, 
are shown in Figure 7. The predicted field star numbers 
were compared with source counts in the magnitude interval $K = 12 - 15$ obtained 
from 2MASS images that sample areas around each cluster. Excellent agreement 
between the observed and predicted counts was found for 
NGC 2401. However, the models overestimate the field star numbers 
for the other two clusters by factors of $1.30\times$ (Haffner 17) and $1.16\times$ 
(NGC 3105). To correct the models for these discrepancies, the predicted 
field star counts for each cluster were scaled over all magnitudes by the same factor to 
force agreement with the 2MASS field star densities. Given 
the lack of checks on the numbers of field stars for 
magnitudes $K \geq 15$ then the field star corrections 
are a source of uncertainty near the faint end of these data. 

	Also shown in Figure 7 are model LFs generated from the Bressan et al. (2012) 
isochrones. The models assume a coeval stellar content with 
a Chabrier (2001) lognormal MF, and a metallicity Z = 0.016. 
The first assumption is consistent with the modest age 
dispersion among stars in low mass (e.g. Hosokawa et al. 
2011; Delgado et al. 2012; Bik et al. 2012) and at least some high mass 
(e.g. Kudryavtseva et al. 2012) clusters. The models have been scaled to match 
the observed number counts in the magnitude intervals indicated in each panel. 
The faint limit used for normalization was set at the same 
M$_K$ for each cluster, whereas the bright limit 
was fixed at $K = 14$, as the number of stars in each cluster plummets 
at brighter magnitudes than this.

	{\bf The artificial star experiments discussed in Section 3 indicate 
that the observations are complete for all clusters when $K < 19.5$, and so sample 
completeness is not an issue for the comparisons made in Figure 7.} 
The LF of Haffner 17 is flat, and is not matched by the models. 
The models fall $\sim 1$ dex above the Haffner 17 LF at the faint end, 
indicating a substantial deficiency in low mass sources when compared 
with the Chabrier MF. The agreement between the models and 
observations would not be improved if a different magnitude interval was 
adopted for scaling the models.

	The difference between the models and the 
observed counts in the upper panel of Figure 7 diminishes towards older ages, and 
better agreement could be obtained near $K = 18$ 
if Haffner 17 were older than estimated here. However, 
the age of Haffner 17 would have to be in error by $\geq 0.1$ Gyr 
to match the number of objects at $K = 18$, and this 
is unlikely given the location of the MSTO in the Pedredos (2000) CMD. Even then, 
a substantial disagreement between the models and 
number counts would remain at fainter magnitudes.

	In contrast to Haffner 17, the model LFs match the NGC 2401 number counts over 
a large part of the LF. As NGC 2401 has the smallest 
fractional field star contamination, then its LF should be the 
least susceptible to uncertainties in field star contamination at the faint end. 
The models match the steady rise in number counts towards fainter magnitudes in NGC 
2401 between $K = 14$ and 16, which is the interval populated by bright MS stars. 
This agreement is consistent with the finding by Baume et al. (2006) that the MF 
of objects with masses $\geq 1$ M$_{\odot}$ in NGC 2401 have a solar neighborhood-like 
MF. The LF of NGC 2401 flattens when $K \geq 16$, due to a change in the 
mass-luminosity relation of MS stars. This break is also reproduced by the 
models, and provides a loose check on the distance modulus. 

	The agreement between the models and the NGC 2401 LF is not so good for 
$K > 16.5$, with neither model providing a satisfactory match to the observations in 
this magnitude range. Better agreement across the entire magnitude range could be 
obtained if the model normalization interval were extended to fainter magnitudes. 
Such a re-normalization would still produce agreement with the 
observations at the bright end within the error bars, although with the models falling 
systematically below the observations.

	As for NGC 3105, there is statistical agreement between the 
models and the LF of this cluster near the bright end. Sagar et al. (2001) find 
that the bright members of NGC 3105 follow a MF with an exponent that is close to 
the Salpeter value, and the results in Figure 7 are consistent with this. 
However, statistically significant differences between the models and the NGC 3105 
number counts occur in Figure 7 when $K \geq 18.5$. A broader normalization interval 
that encompasses fainter magnitudes would improve the agreement at the faint end 
slightly, although the quality of the fit at the bright end would then diminish. 

	In Section 4 it was demonstrated that the distance modulus 
computed from the $(K, J-K)$ CMD depends on the metallicity of the isochrone that 
is compared with the CMD, and this could affect the agreement between the LFs 
and the models. NGC 3105 is the cluster in the present sample that is most likely to 
have a sub-solar metallicity given its large R$_{GC}$. The green curves 
in the lower panel of Figure 7 show the result of adopting a distance modulus that is 
0.15 dex smaller (i.e. corresponding to what would be obtained from 
an isochrone with a metallicity that is 1/2 Solar -- Section 4) for NGC 3105. 
The agreement between the observations and the models does not change greatly 
when this lower distance modulus is applied. The conclusions reached in this section 
are thus not sensitive to uncertainties in the adopted distance moduli.

	In addition to providing insight into the MF, the LF of 
a star cluster also contains information about its age. 
The trajectory of the PMS sequences in Figures 1 - 3 tend to flatten near the MSCO, 
resulting in a build-up of PMS objects in a narrow magnitude interval, creating a bump 
in the LF that can be used as an age indicator (e.g. Cignoni et al. 2010; Davidge 2012). 
The ability to detect this feature depends on factors such as age, the number 
of stars, and the binning used to construct the LFs. Cignoni et al. 
(2010) argue that this feature is a useful age indicator for well-populated systems 
that are younger than $\sim 30$ Myr. 

	Can the LFs in Figure 7 be used to place constraints on the cluster ages? 
Haffner 17 is not considered when answering this question given the difficulty matching 
the overall shape of the cluster LF. As for NGC 2401, if it has an age 
log($t_{yr}$) = 7.4 then the model predicts that there should be a PMS bump near $K = 
17$, while for an age log($t_{yr}$) = 7.6 then the bump is 0.5 magnitudes fainter. 
The NGC 2401 LF does not contain a feature at either magnitude, although the 
amplitudes of the bumps in the models are comparable to 
the error bars in the LF. While the dip in number counts 
between $K = 16.5$ and $K = 17$ might be consistent with an age 
log($t_{yr}) \sim 7.2$, the statistical significance of this feature is low. 
There are no obvious signatures of a PMS-related bump in the NGC 3105 LF. These
comparisons suggest that the PMS bump is only a useful age indicator for systems 
that are more massive than NGC 2401 and NGC 3105 (i.e. masses 
$> 5 \times 10^2$M$_{\odot}$. 

\section{THE CORRELATION PROPERTIES OF CLUSTER STARS}

	The spatial distribution of stars in a cluster evolves with time as 
the system relaxes. However, non-secular processes may also affect the spatial 
distribution of cluster stars. For example, the expulsion of large quantities of gas 
early in the life of a system due to winds and/or supernovae (SNe) 
will change the gravitational potential. The system will expand 
(e.g. Baumgardt \& Kroupa 2007), and stars with velocities that exceed 
the new (lower) escape velocity of the mass-depleted system will be lost.

	The clustering properties of three samples of objects in Haffner 17, 
NGC 2401, and NGC 3105 are examined: (1) all detected objects (the `ALL' sample), 
(2) bright main sequence stars (the `BMS' sample), which have 
M$_K \leq 2.5$ and are on the MS, and (3) faint main sequence/ pre-main sequence 
objects (the `FMS' sample), which have M$_K$ between 3 and 5. Absolute magnitudes 
were computed using the distance moduli obtained in Section 4. The degree of 
clustering is measured with the two-point angular correlation function (TPCF). 
The core component of the TPCF is the separation function (SF), which is the 
distribution of separations between all possible source pairings. 
The SF multiplexes information from all objects, and so makes efficient 
use of available information, although at the price of mixing information 
from cluster and non-cluster members. Geometric effects are rectified by 
dividing the SF by that of an artificially-generated 
uniformly-distributed sample of objects with an image geometry 
that mimics the actual data. This ratio is then normalized 
according to the number of pairings in the two SFs.

	The TPCFs of the ALL samples are compared in the top panel of 
Figure 8. The TPCFs reflect the clustering properties of a mix of 
cluster members and field stars, and the impact on field star 
contamination on the TPCFs is examined later in this section.
Still, the TPCFs of the cluster ALL samples are similar over most separations, 
with uniform clustering properties for separations between 20 and 40 arcsec. The 
TPCFs of the ALL samples diverge at separations $> 76 - 80$ arcsec. 

	The TPCFs of the ALL samples differ at small separations, and this 
may signal differences in the binary star contents. Binaries in nearby 
star-forming regions have separations $< 0.04$ pc (Larson 1995), and the widest 
such binaries are resolved with these data in all three clusters. However, 
cluster-to-cluster comparisons are complicated by differences 
in the distance and angular resolution. Consider Haffner 
17, where there is a peak in the TPCF at the smallest separations in the upper 
panel of Figure 8. The bin that covers the smallest separations samples 
objects with spatial separations from 0.0024 pc (i.e. 0.12 arcsec -- the angular 
resolving power of the data) to 0.04 pc (i.e. 2 arcsec - the bin size used to construct 
the TPCF). The bin that samples the smallest separations in Haffner 17 thus covers 
a clean population of binaries. In contrast, NGC 3105 is the most distant cluster, 
and the smallest separation bin in that cluster samples objects with 
projected separations of 0.0063 -- 0.070 pc. Therefore, 
binaries at spatial separations that would be detected in Haffner 17 
are not resolved in NGC 3105, and the binary signal in NGC 3105 is 
diluted by non-binaries in the smallest separation bin.

	The clustering properties of stars having different masses -- 
such as the objects in the BMS and FMS samples -- provide 
insights into the dynamical state of a system. 
For example, objects with lower masses being more uniformly distributed on 
the sky than higher mass objects is a signature 
of mass segregation, although other processes could produce similar results. 
The ratio of the SFs of the BMS and FMS samples are shown in the lower panel of 
Figure 8. If the objects in the BMS and FMS samples have 
similar clustering properties then the ratio of the SFs 
should be $\sim 1$, and this is the case over most separations in NGC 2401. 
Significantly, the LF of NGC 2401 was also found to be similar to the 
solar neighborhood (Section 5). If it is assumed that the MF is 
universal then this suggests that mass segregation is not yet 
well advanced in the mass range probed in NGC 2401.

	The situation is different for Haffner 17 and NGC 3105. While the 
ratio of the BMS and FMS SFs is close to unity 
in both clusters for separations $< 60$ arcsec, this is not the case 
at higher separations. This behaviour indicates that sources in the FMS samples 
are more uniformly distributed across the GSAOI field over large angular 
scales than the objects in the BMS samples. However, field star contamination is 
substantial in both clusters, and this must be considered when interpreting these 
results.

	Field stars are uniformly distributed on the sky and will bias TPCFs 
towards flatter distributions. If the magnitude intervals that cover the BMS 
and FMS samples have different fractional contaminations from field stars 
then the ratio of the BMS and FMS SFs will be skewed by these 
differences. The field star fraction in the Haffner 17 and NGC 3105 data
is higher in the FMS sample (70\% in Haffner 17 and 64\% in NGC 
3105) than in the BMS sample (29\% in Haffner 17 and 41\% in NGC 3105). For comparison, 
the field star fraction in the NGC 2401 BMS and FMS samples -- where the objects 
show similar clustering properties (see above) -- are not greatly 
different, with 26\% contamination in the BMS sample, and 29\% in the FMS sample.

	It is difficult to separate cluster and field stars 
in the absence of proper motion and radial velocity measurements. However, 
the role that field stars play in the BMS and FMS samples can be assessed in 
a statistical manner. Following the procedure described by Davidge et al. (2013), 
objects that have a uniform on-sky distribution -- thereby simulating the distribution 
of field stars -- were added to the BMS samples of Haffner 17 and NGC 3105 
to boost artificially the field star fraction to match that in the 
corresponding FMS samples. A number of realizations were run, and a mean BMS$+$field 
star SF was computed for both clusters to suppress stochastic variations. 
The results are compared with the observed BMS TPCFs in the top panel of Figure 9.

	Introducing a simulated field star population has only a modest impact on 
the BMS TPCFs at small separations, reflecting the modest 
on-sky density of the added stars, and the resulting low 
probablity that they are near other objects. However, the multiplexed 
nature of the SFs causes the contribution made by the 
added stars to grow towards progressively larger separations. It can be seen in Figure 9 
that balancing the field star fractions affects the TPCFs at 
separations $> 60$ arcsec, which is where differences between the BMS and FMS samples
of Haffner 17 and NGC 3105 in Figure 8 are apparent.

	The ratios of the mean BMS$+$field star and FMS TPCFs of Haffner 17 and 
NGC 3105 are shown in the lower panel of Figure 9. Balancing the field star levels in 
the BMS and FMS samples moves the ratio of TPCFs at separations 
$> 60$ arcsec in Haffner 17 much closer to unity, as expected if the 
differences in Figure 8 between the Haffner 17 samples are significantly skewed by field 
stars. Still, it is apparent that the BMS and FMS samples in this cluster 
have different angular distributions, as the BMS/FMS ratio falls consistently below 
unity at separations $> 80$ arcsec after balancing the field star fractions. As for 
NGC 3105, the FMS sample also remains more uniformly distributed than the BMS component 
after balancing the field star fractions. There are thus mass-related differences 
in the spatial distribution of stars in Haffner 17 and NGC 3105. This is evidence 
that these clusters are dynamically evolved.

\section{DISCUSSION \& SUMMARY}

	Near-infrared images obtained with GSAOI$+$GeMS have been used to 
examine the photometric properties and spatial distributions of stars in 
the young open clusters Haffner 17, NGC 2401, and NGC 3105. 
These clusters are located at large Galactocentric radii, and NGC 3105 is one 
of the most remote young clusters yet discovered in the Galaxy. Studies of external 
galaxies suggest that young and intermediate age clusters are not uncommon 
in the outermost regions of star-forming disks (e.g. Herbert-Fort et al. 2012).
Studies of young clusters in the outer Galaxy may then yield insights into how 
clusters in the outer disks of spiral galaxies evolve in general. Because these clusters 
are in an environment that is systematically different from the Solar Neighborhood, 
their evolution might differ from those of better-studied nearby clusters.

	The age, reddening, and distance modulus estimated for each cluster 
from the GSAOI$+$GeMS data are summarized in Table 3. 
The distances and reddenings found here are based on the morphology of the MS 
on the $(K, J-K)$ CMD. The distances measured for NGC 3105 and NGC 2401 place 
them closer than previous studies, while the opposite is true for Haffner 17. 
The reddenings of all three clusters also differ from previous estimates. It should be 
recalled that the $E(B-V)$ values found here were computed from $E(J-K)$ 
using the entries in Table 6 of Schlegel et al. (1998). If the extinction along the 
lines of sight to the target clusters follows different reddening laws 
then this will affect comparisons with $E(B-V)$ estimates made at visible wavelengths.

	The ages are based on the photometric properties of PMS and lower mass MS 
stars, and so are independent of previous age estimates that relied on 
the MSTO. Barring a peculiar MF, ages that rely on faint cluster members will be less 
susceptible to the statistical variations in stellar content that can affect the 
brightest MS and evolved stars in low mass clusters, which in turn complicates efforts 
to determine ages from the MSTO. This being said, the use of faint objects 
to measure ages comes at a cost. Uncertainties in the photometric measurements increase 
towards fainter magnitudes, and sample completeness may become a factor when asessing 
the magnitude of the MSCO. Contamination from field stars also becomes a concern when 
measuring ages from faint members, as the number of field stars grows towards fainter 
magnitudes. The predominantly red colors of field stars also mean that they will tend 
to overlap more with those of PMS stars than with stars near the MSTO. Finally, there 
are also numerous uncertainties in models of PMS objects (e.g. Seiss 2001). Still, it is 
encouraging that the isochrones used here track the observed PMS sequences, 
and yield ages that are consistent with previous determinations. 

	Haffner 17 has the most sparsely populated CMD in our sample, and so 
it might be aniticipated that the age measured from the MSTO of that cluster will 
be the most prone to statistical variations among the brightest MS stars. However, 
there is no evidence that this is the case, as the age determined from the GSAOI data 
is consistent with that found by Pedredos (2000). As for NGC 2401 
and NGC 3105, the GSAOI data suggest that these clusters have 
ages 20 -- 30 Myr. Previous studies of NGC 2401 have suggested a 
wide range of ages, and our results favor the age found 
by Baume et al. (2006), as opposed to those measured by Sujatha et al. (2004) and 
Hasegawa et al. (2008). The age measured for NGC 3105 agrees with those deduced by 
Sagar et al. (2001) and Paunzen et al. (2005) from the MSTO. The good agreement 
with published ages suggests that stochastic effects have not affected the present-day 
MSTO contents of these clusters.

	Given their ages, the clusters are expected to contain 
faint sub-solar mass PMS objects, and PMS sequences are seen in the CMDs of NGC 
2401 and NGC 3105. An intriguing result from Section 4 is that the mean 
Br$\gamma - K$ color changes within $\sim 0.6$ magnitudes in $K$ 
of the onset of the PMS sequence in both clusters. In addition, 
the photometric properties of stars along the PMS sequence in NGC 3105 differ 
from those of field stars in the same magnitude interval, in the sense that the PMS 
stars are brighter on average in the Br$\gamma$ filter at a given $K$ magnitude than 
field stars. While narrow-band Br$\gamma$ observations alone do not provide the 
information required for an unambiguous interpretation of the cause of these differences, 
both results are consistent with Br$\gamma$ emission being present. Optical 
and/or near-infrared spectroscopy will be required to determine if line emission 
is present. If Br$\gamma$ emission is present among PMS stars in NGC 2401 and NGC 
3105 then it must have a high frequency of occurrence to affect the mean 
Br$\gamma - K$ colors. 

	Star clusters disperse in response to internal and external processes. 
Despite having more-or-less similar ages, our data suggest that Haffner 
17, NGC 2401, and NGC 3105 are in different stages of evolution. 
The $K$ LFs discussed in Section 5 provide some of the evidence that supports 
this interpretation. The $K$ LF of NGC 2401 is well-matched over a wide magnitude 
range by models that assume a Chabrier (2001) MF. To the extent that 
the initial MF of stars having masses within a few tenths of a dex of solar 
is universal, then this suggests that NGC 2401 has not shed large numbers of low mass 
stars. However, this is not the case for NGC 3105, as the $K$ LF of this cluster 
is deficient in low mass objects when compared with models. 
The $K$ LF of Haffner 17 is very different from those of either NGC 2401 and 
NGC 3105, in that it is relatively flat, even at the bright end.

	The spatial distributions of MS and PMS stars provide 
additional evidence that the clusters are in different stages of evolution. 
BMS and FMS objects in NGC 2401 have similar clustering properties, suggesting 
that intermediate and low mass objects in NGC 2401 have not yet developed 
kinematically distinct distributions, and this is further evidence that 
NGC 2401 is not dynamically evolved. In contrast, after 
balancing field star contamination in different magnitude intervals, 
PMS objects and low mass MS stars in Haffner 17 and NGC 3105 
are more uniformly distributed than stars in the BMS sample, suggesting 
that low and high mass objects in these clusters have different kinematic properties.

	Insights into the evolutionary status of these clusters can also be 
deduced from their integrated properties. Cluster masses, radii, and free-fall times 
were calculated for each cluster, and the results are shown in Table 3. 
The masses were computed by transforming the cluster LFs into MFs using 
magnitude $vs$ mass relations from the Padova isochrones, and then integrating 
the results. These masses are lower limits because (1) the clusters 
extend beyond the GSAOI field, and (2) cluster members with masses $\leq 0.3$ 
M$_{\odot}$ are not sampled with these data. Still, the corrections for 
these effects are likely to be modest. An alternative would be to compute 
total masses from the integrated brightnesses listed in Table 1, although the results 
would be highly uncertain due to the presence of bright stars (Section 1). 

	The last column of Table 3 lists the ratio of the cluster age to the dynamical 
time scale ($\tau_{dyn}$), which is one measure of the kinematic state of a system. 
Marks \& Kroupa (2012) find that significant evolution in cluster density occurs when a 
system is younger than $\sim 10 \tau_{dyn}$, and that only more subdued changes in 
density occur after this time. To the extent that the ratio of cluster age to the 
dynamical timescale is a simple probe of kinematic evolution, it indicates 
that NGC 2401 is the least evolved cluster (i.e. it has the smallest ratio of 
age to $\tau_{dyn}$). This is consistent with the investigation of the MF 
and the angular distributions of BMS and FMS stars presented above. In contrast, 
Haffner 17 and NGC 3105 are predicted to be more evolved than NGC 2401. 

	If these clusters formed in large-scale complexes then they may be 
susceptible to environmentally-driven evolution. However, 
an examination of Digital Sky Survey images shows that the clusters are 
isolated, in the sense that neighboring stellar groupings are at least 
10 arcmin away in projection. Of course, over tens 
of Myr the random motions that clusters attain in disks may cause them 
to move large distances and erase structure that was in place during the time of 
their formation (e.g. Davidge 2007). Their present-day environment may 
then be very different from the environment where they formed. 

	The evolution and present-day appearance of a cluster can also 
be shaped early in its life by internally-driven events that 
can cause significant cluster-to-cluster differences in stellar content. 
Systems that form massive, short-lived stars may be prone 
to a catastrophic loss of gas driven by winds and SNe that 
can change the gravitational potential and cause the cluster to disperse 
(e.g. Lada \& Lada 2003). If SNe activity is delayed then the likelihood of a cluster 
being disrupted will drop, as the fraction of cluster 
mass that is in gas and dust decreases with time as this material is consumed by star 
formation and disperses from the cluster -- as time passes the removal of gas will 
then have a smaller impact on the cluster potential. Any delay in SNe activity 
also allows cluster members to come closer to being fully virialized, with the result 
that  the extremes in velocity that are conducive to the loss of stars when the 
cluster potential changes are reduced (e.g. Smith et al. 2011). 

	The total stellar mass of a cluster likely plays a key 
role in determining if internal processes drive its evolution. 
Pelupessy \& Portegies Zwart (2012) investigate the role that statistical 
variations among massive stars play in the evolution of systems 
with stellar masses $\sim 350$ M$_{\odot}$. The three 
clusters studied here have masses close to that adopted for the simulations, allowing 
relevant comparisons to be made. Pelupessy \& Portegies Zwart find that stochastic 
variations in the population of upper MS stars can lead to profound cluster-to-cluster 
differences in the early evolution of the simulated clusters. 
If stars are present that are massive enough to 
become SN early-on, then low mass stars -- which attain higher velocities than more 
massive stars as a system relaxes -- may be preferentially lost from the cluster early 
in its evolution. On the other hand, if statistical flucuations 
in stellar content are such that very massive stars do not 
form then a cluster may not be purged of gas until much later 
in its life, with a smaller impact on its structural characteristics. 
Of course, the timing of SNe activity is not the only factor that shapes early cluster 
evolution. For example, sub-clustering may concentrate star-forming activity, lowering 
local gas fractions and thus reducing the impact of SNe-driven gas loss 
(Kruijssen et al. 2012). 

	The properties of the NGC 2401 LF and the 
relative distributions of its stars in the BMS and FMS samples hint 
that NGC 2401 was not subjected to violent early evolution, and had 
an uneventful past. However, the same can not be said for Haffner 17 and NGC 3105, 
and there are observations that could be carried out that would cast light on their early 
histories. Clusters will expand in response to the sudden loss of large quantities 
of gas (e.g. Baumgardt \& Kroupa 2007), and an extended stellar halo might be an 
observational signature of this (e.g. Davidge 2012; Bastian \& 
Goodwin 2006). If Haffner 17 and NGC 3105 experienced early mass loss then a wide-field 
survey of these clusters may reveal a halo containing objects that span 
a range of masses, although the long term survivability of such a structure will 
depend on the local tidal field. In the case of NGC 3105 some of the 
halo objects should be PMS sources, and these may not be difficult to detect if they are 
hydrogen line emitters, as suggested in Section 4. 

	In addition to affecting the structural properties of star clusters, 
the presence of massive hot stars may also affect the final MF of stars that form. 
The ultraviolet radiation field produced by very massive stars may 
erode nearby accretion disks and choke subsequent stellar growth, thereby affecting 
the minimum stellar mass in their immediate vicinity (e.g. Johnstone 
et al. 1998; Adams et al. 2004; De Marchi et al. 2011). If Haffner 17 contained an 
extremely massive hot star then such an object may have affected the final cluster MF.

	The preceeding discussion assumes that the processes that drive the disruption 
of a cluster during the initial stages of its assembly differ from those that 
disrupt older clusters. However, there are hints that the disruption rate of 
very young clusters may be similar to that of older clusters, as would be expected if 
cluster destruction during the early stages of cluster assembly and at later times 
are driven by the same mechanisms. In particular, if all stars form in 
clusters, then the relative numbers of stars that are in clusters 
and the field in the Solar neighborhood suggest that over 90\% of natal groupings 
are disrupted within 10 Myr of their birth (Bonatto \& Bica 2011). If it is assumed that 
clusters remain intact over the lifetimes of their most massive stars (i.e. $\sim 1$ 
Myr), then this pace of attrition is consistent with that seen among clusters that 
span a broad range of ages in nearby disk galaxies, where the number of clusters drops 
by roughly an order of magnitude for every decade in age (Fall \& Chandar 2012). That 
the pace of cluster destruction among very young systems tracks that among much older 
systems is suggestive of similar mechanisms disrupting clusters over a wide range of 
ages. If this is the case then while very young clusters are undoubtedly subject to mass 
loss driven by winds and SNe, these may not be the dominant mechanisms driving 
cluster disruption.

\acknowledgements{Thanks are extended to the anonymous referee for providing a prompt and 
helpful report.}
 
\parindent=0.0cm

\clearpage

\begin{table*}
\begin{center}
\begin{tabular}{lcccccccc}
\tableline\tableline
Cluster & $\ell$ & $b$ & log(age)\tablenotemark{1} & $\mu_0$\tablenotemark{1} & $K_{Tot}$\tablenotemark{2} & Radius\tablenotemark{2} & M$_K^{Tot}$ \\
 & (Degrees) & (Degrees) & (years) & & & (arcsec) & \\ 
\tableline
Haffner 17 & 247.7 & -2.5 & 7.7 & 12.3 & 7.3 & 75 & --5.0 \\
NGC 2401 & 229.7 & 1.9 & 7.4 & 14.0 & 10.7 & 45 & --3.3 \\
NGC 3105 & 279.9 & 0.3 & 7.2 & 13.4 & 5.6 & 80 & --7.8 \\
\tableline
\end{tabular}
\tablenotetext{1}{From August 2013 version of WEBDA database.}
\tablenotetext{2}{Measured from 2MASS images.}
\caption{Cluster Parameters}
\end{center}
\end{table*}

\clearpage

\begin{table*}
\begin{center}
\begin{tabular}{lcccllcc}
\tableline\tableline
Cluster & Date & $r_0$\tablenotemark{1} & $<R>$\tablenotemark{2} & Filter & Exposure Times & $<FWHM>$\tablenotemark{3} & $\sigma_{FWHM}$\tablenotemark{4} \\
 & Observed & (cm) & & (sec) & (arcsec) & (arcsec) \\
\tableline
Haffner 17 & Jan 29, 2013 & 13.0 & 13.8 & J & $3 \times 10; 5 \times 25$ & 0.119 & $\pm 0.012$ \\
 & & & & Ks & $3 \times 10; 5 \times 25$ & 0.098 & $\pm 0.005$ \\
 & & & & Br$\gamma$ & $3 \times 10; 5 \times 60$ & -- & -- \\
 & & & & & & & \\
NGC 2401 & Jan 30, 2013 & 11.0 & 14.3 & J & $3 \times 10; 5 \times 40$ & 0.093 & $\pm 0.010$ \\
 & & & & Ks & $3 \times 5; 5 \times 25$ & 0.083 & $\pm 0.007$ \\
 & & & & Br$\gamma$ & $3 \times 5; 5 \times 60$ & -- & -- \\
 & & & & & & & \\
NGC 3105 & Jan 26, 2013 & 9.0 & 11.5 & J & $3 \times 10; 5 \times 40$ & 0.178 & $\pm 0.017$ \\
 & & & & Ks & $3 \times 10; 5 \times 40$ & 0.117 & $\pm 0.012$ \\
 & & & & Br$\gamma$ & $3 \times 10; 5 \times 60$ & -- & -- \\
 & & & & & & \\
\tableline
\end{tabular}
\tablenotetext{1}{Fried parameter at $\lambda = 500$ nm measured at the mid-point of the cluster observing sequence.}
\tablenotetext{2}{Mean R magnitude of NGSs.}
\tablenotetext{3}{Mean FWHM measured from the PSF stars in the long exposure images.}
\tablenotetext{4}{Standard deviation about $<FWHM>$.}
\caption{Details of the Observations}
\end{center}
\end{table*}

\clearpage

\begin{table*}
\begin{center}
\begin{tabular}{lccccccc}
\tableline\tableline
Cluster & Age\tablenotemark{1} & E(B-V)\tablenotemark{1} & $\mu_0$\tablenotemark{1} & Mass & Radius & $\tau_{dyn}$\tablenotemark{2} & Age/$\tau_{dyn}$ \\
 & (Myr) & & & (M$_{\odot}$) & (pc) & (Myr) & \\
\tableline
Haffner 17 & 52 & 0.85 & 13.0 & 270 & 1.7 & 3.2 & 16 \\
NGC 2401 & 25 & 0.15 & 13.7 & 230 & 1.2 & 2.0 & 12 \\
NGC 3105 & 25 & 0.70 & 14.3 & 490 & 2.8 & 1.5 & 17 \\
\tableline
\end{tabular}
\tablenotetext{1}{Values found in this study. The ages quoted for Haffner 17 and NGC 2401 are the midpoints of the age ranges estimated in Section 4.2.2.}
\tablenotetext{2}{Dynamical timescale.}
\caption{Cluster Properties}
\end{center}
\end{table*}

\clearpage

\begin{figure}
\figurenum{1}
\epsscale{1.00}
\plotone{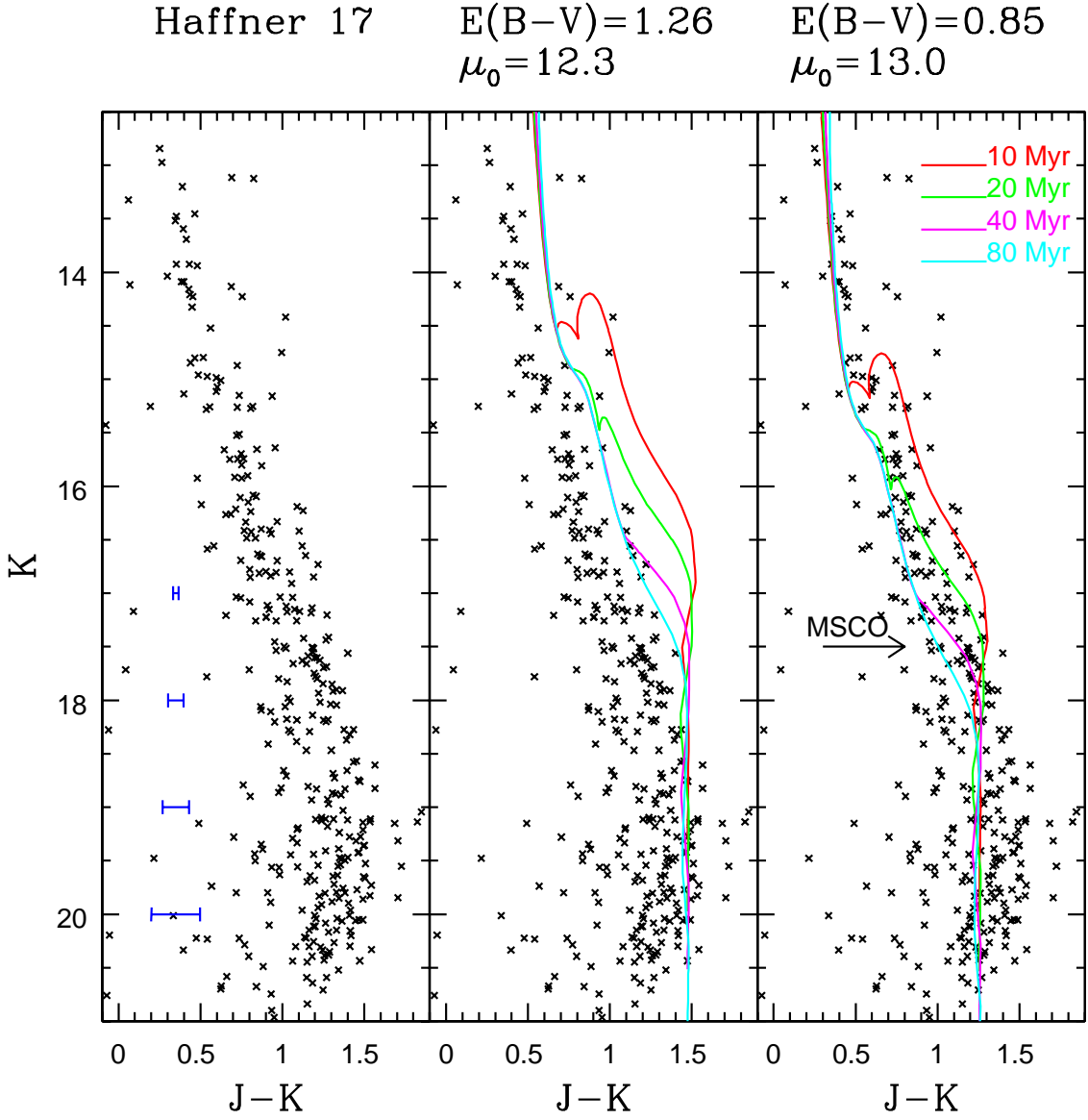}
\caption{The $(K, J-K)$ CMD of Haffner 17 is compared 
with $Z = 0.016$ isochrones from Bressan et al. (2012). {\bf The blue error bars 
in the left hand panel are the $\pm 1\sigma$ random uncertainties calculated from 
artificial star experiments.} The comparisons in the middle panel 
use the reddening and distance modulus determined by 
Pedredos (2000), while those in the right hand panel use the reddening and 
distance modulus found here. The approximate location of the MS cut-off 
(MSCO), determined in Section 4.2, is indicated in the right hand panel.}
\end{figure}

\clearpage

\begin{figure}
\figurenum{2}
\epsscale{1.00}
\plotone{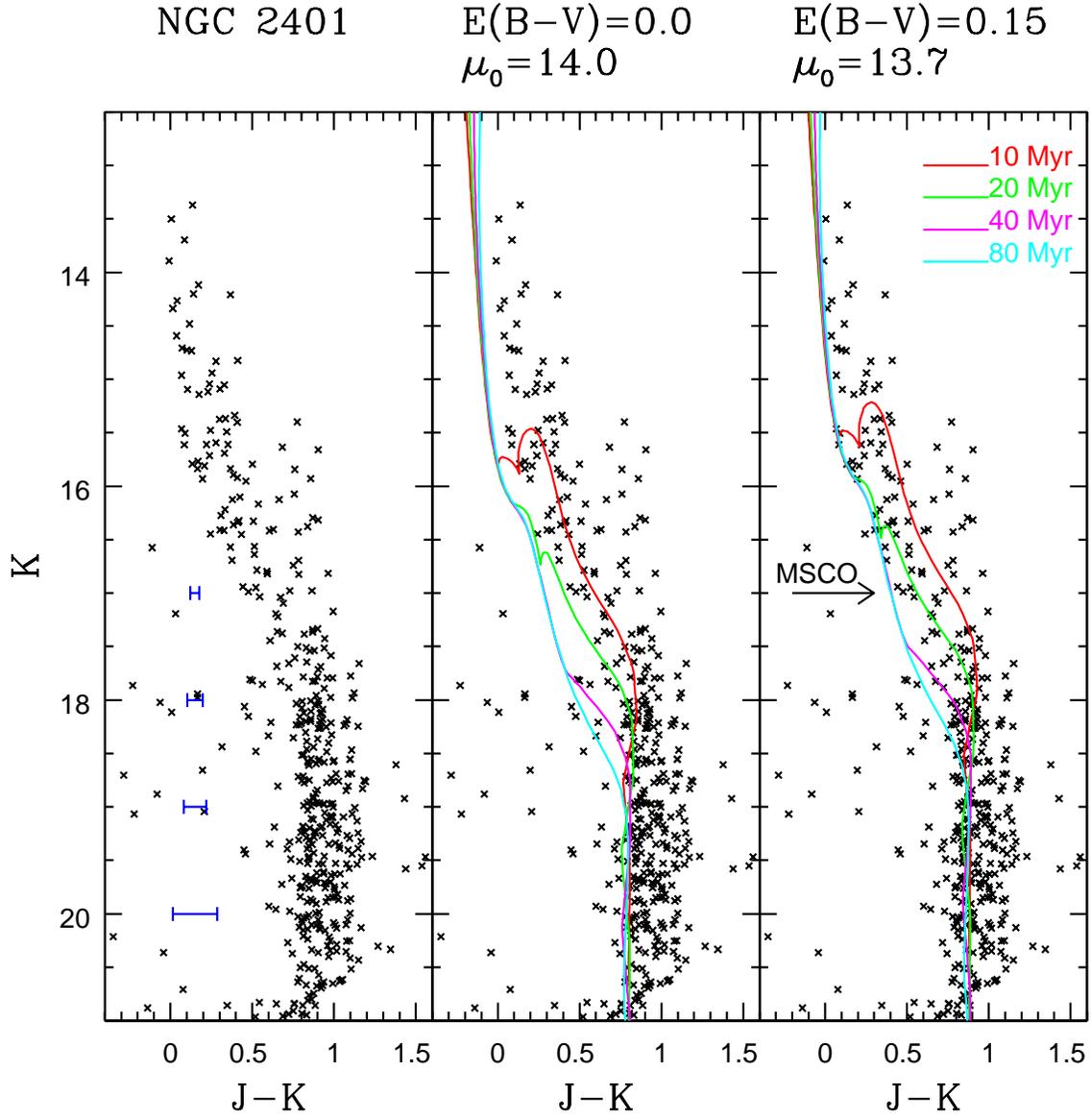}
\caption{The same as Figure 1, but for NGC 2401. The reddening and 
distance modulus used in the middle panel are from Baume et al. (2006).}
\end{figure}

\clearpage

\begin{figure}
\figurenum{3}
\epsscale{1.00}
\plotone{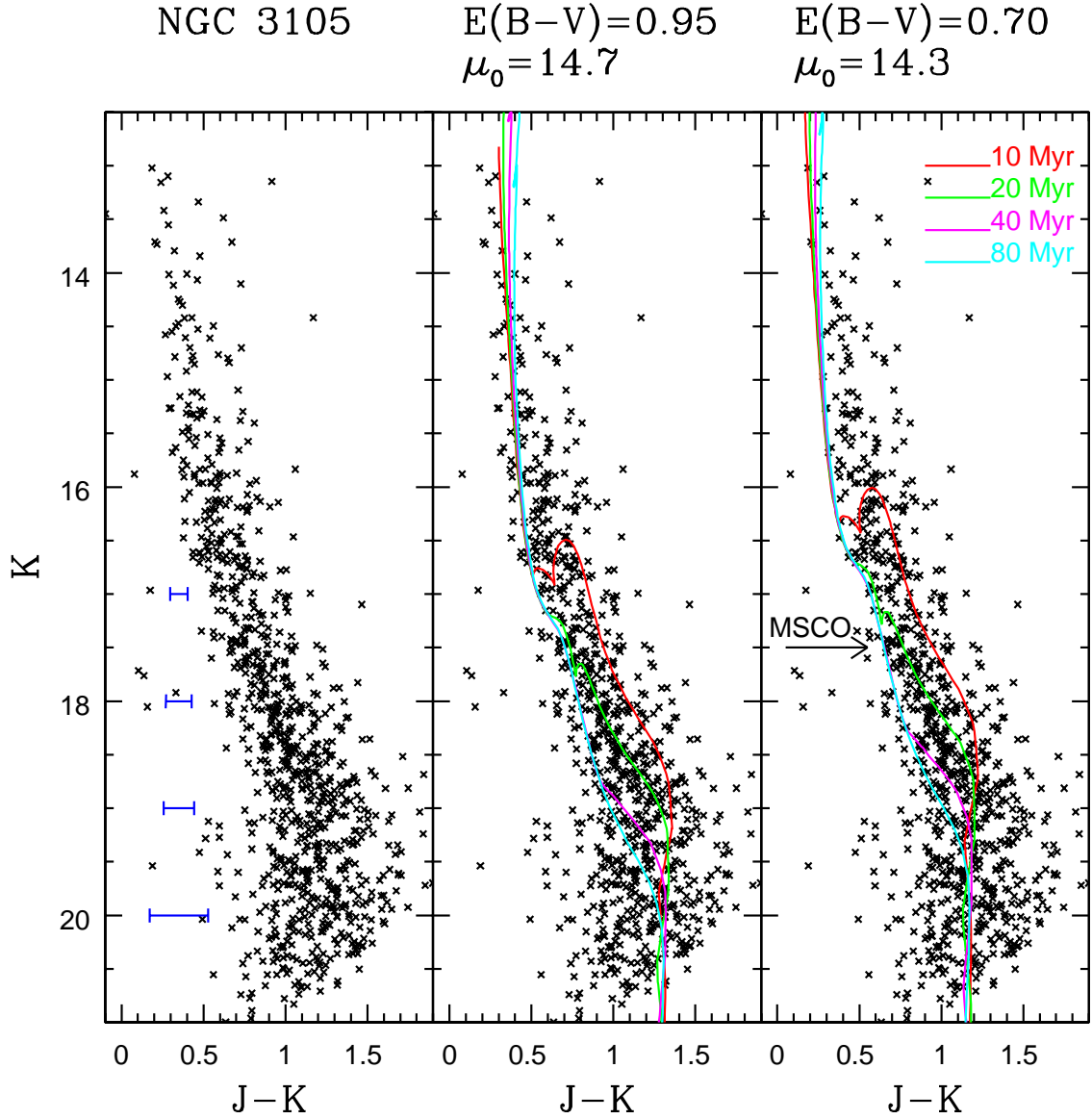}
\caption{The same as Figure 1, but for NGC 3105. The reddening and 
distance modulus used in the middle panel are from Paunzen et al. (2005).}
\end{figure}

\clearpage

\begin{figure}
\figurenum{4}
\epsscale{1.00}
\plotone{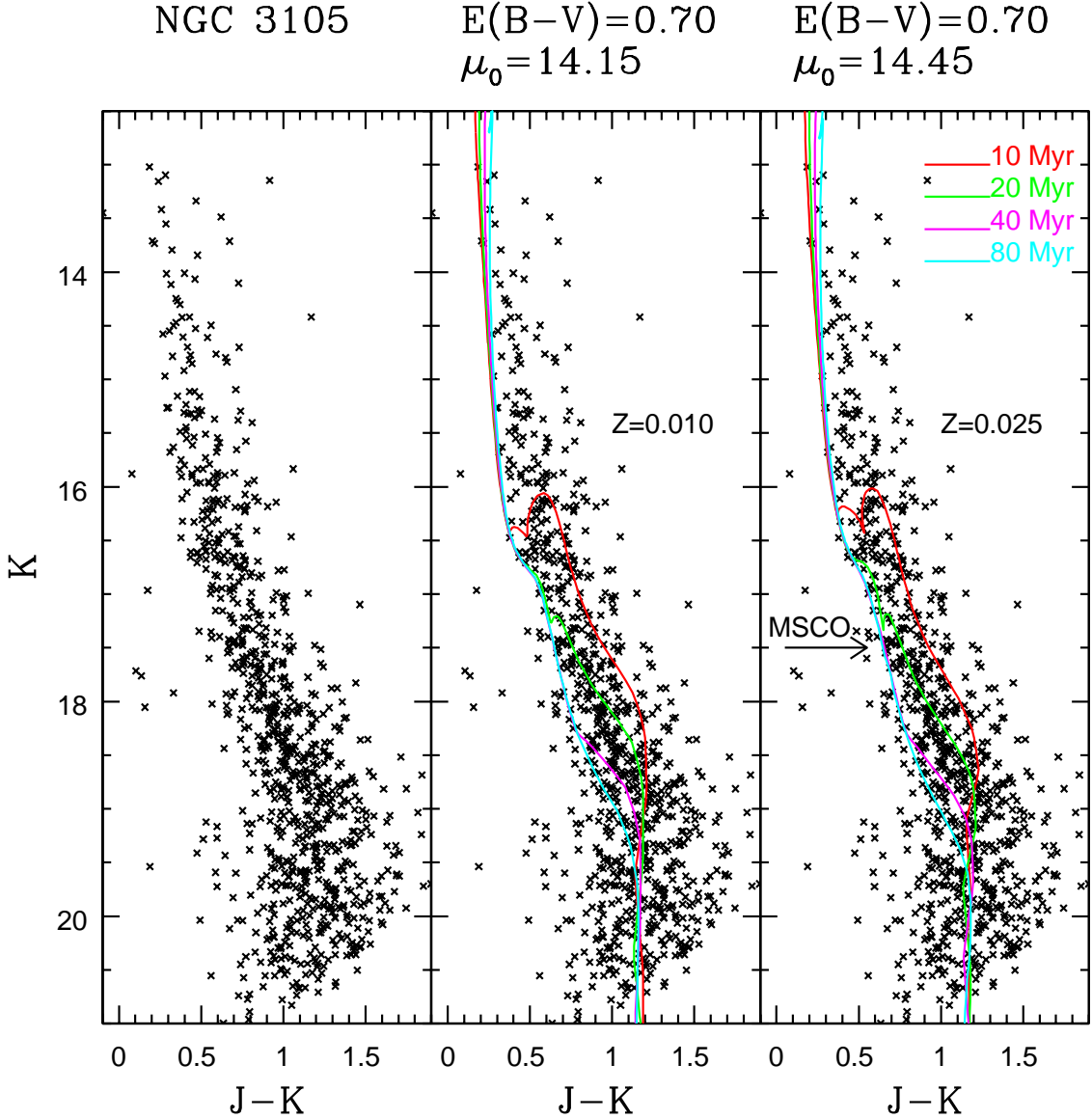}
\caption{The sensitivity of the measured distance and reddening to 
metallicity. The $(K, J-K)$ CMD of NGC 3105 is compared with Z = 0.010 and 
Z = 0.025 isochrones in the middle and right hand panels. $E(B-V)$ and 
$\mu_0$ have been estimated for both metallicities using the criteria 
employed for the comparison in the right hand panel of Figure 3. 
The same reddening is found for both metallicities. However, 
$\mu_0$ differs by 0.30 dex, in the sense of larger $\mu_0$ values as 
metallicity increases. Significantly, the PMS sequence that runs from $K = 
17.5$ to 19 is paralled by the 20 Myr isochrones for both metallicities, 
and is consistent with an age $\sim 25$ Myr.}
\end{figure}

\clearpage

\begin{figure}
\figurenum{5}
\epsscale{1.00}
\plotone{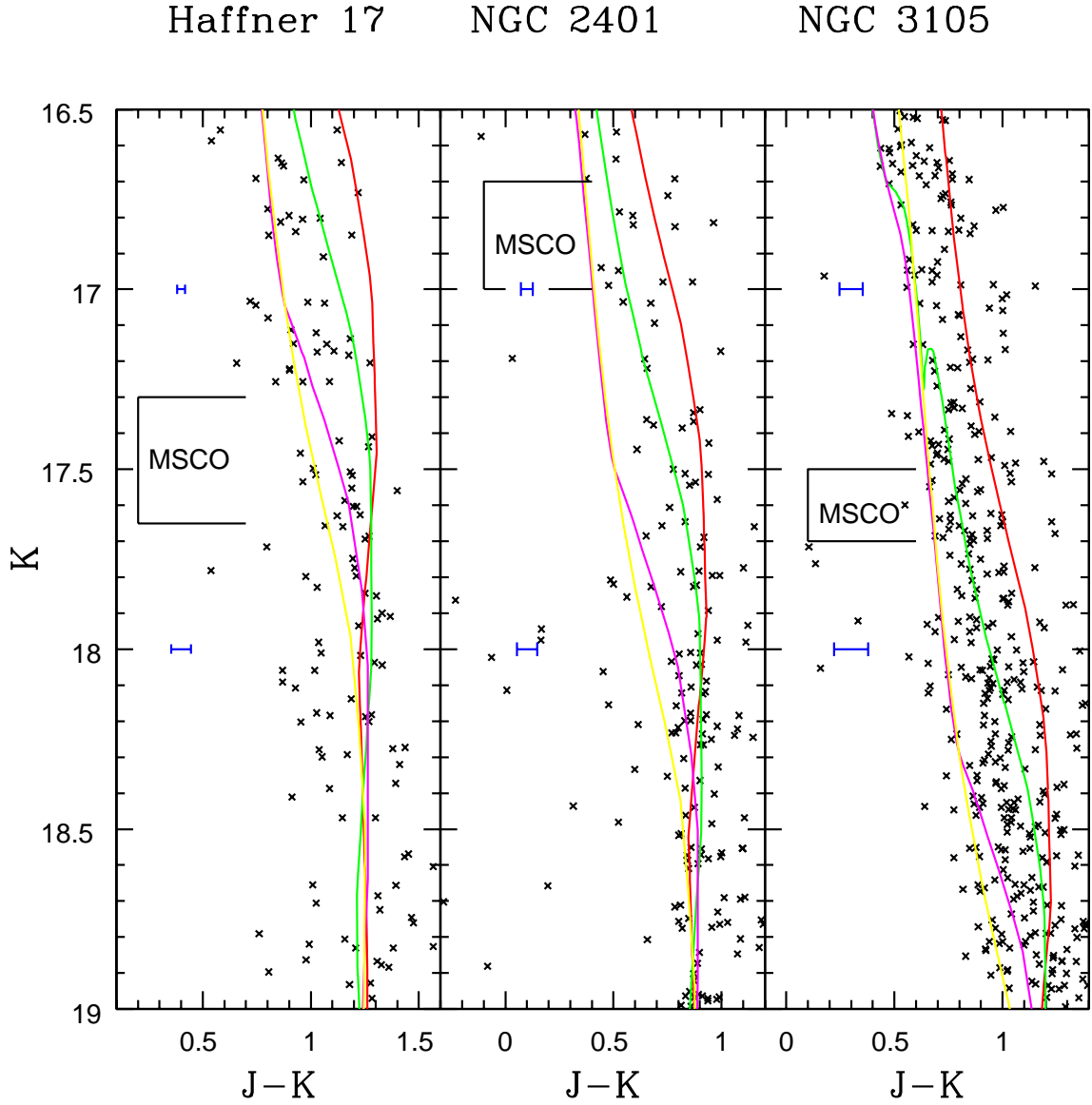}
\caption{The sections of the CMDs containing the lower portions of the MS and 
the upper portions of the PMS. {\bf The blue error bars show the $\pm 1\sigma$ random 
uncertainties in $J-K$ calculated from the artificial star experiments.} Bright and faint 
limits for the MSCO are indicated for each cluster, and the criteria for selecting these 
are discussed in the text. The red, green, and violet lines are isochrones with 
the same color codes used in Figures 1 -- 4, while the yellow sequence is a 
1 Gyr isochrone with $Z = 0.016$. The CMDs of NGC 2401 and 
NGC 3105 contain sequences that are populated by PMS stars. 
The PMS sequence of NGC 2401 is well matched by the 20 Myr isochrone, whereas 
the PMS locus of NGC 3105 is consistent with an age 25 Myr.}
\end{figure}

\clearpage

\begin{figure}
\figurenum{6}
\epsscale{1.00}
\plotone{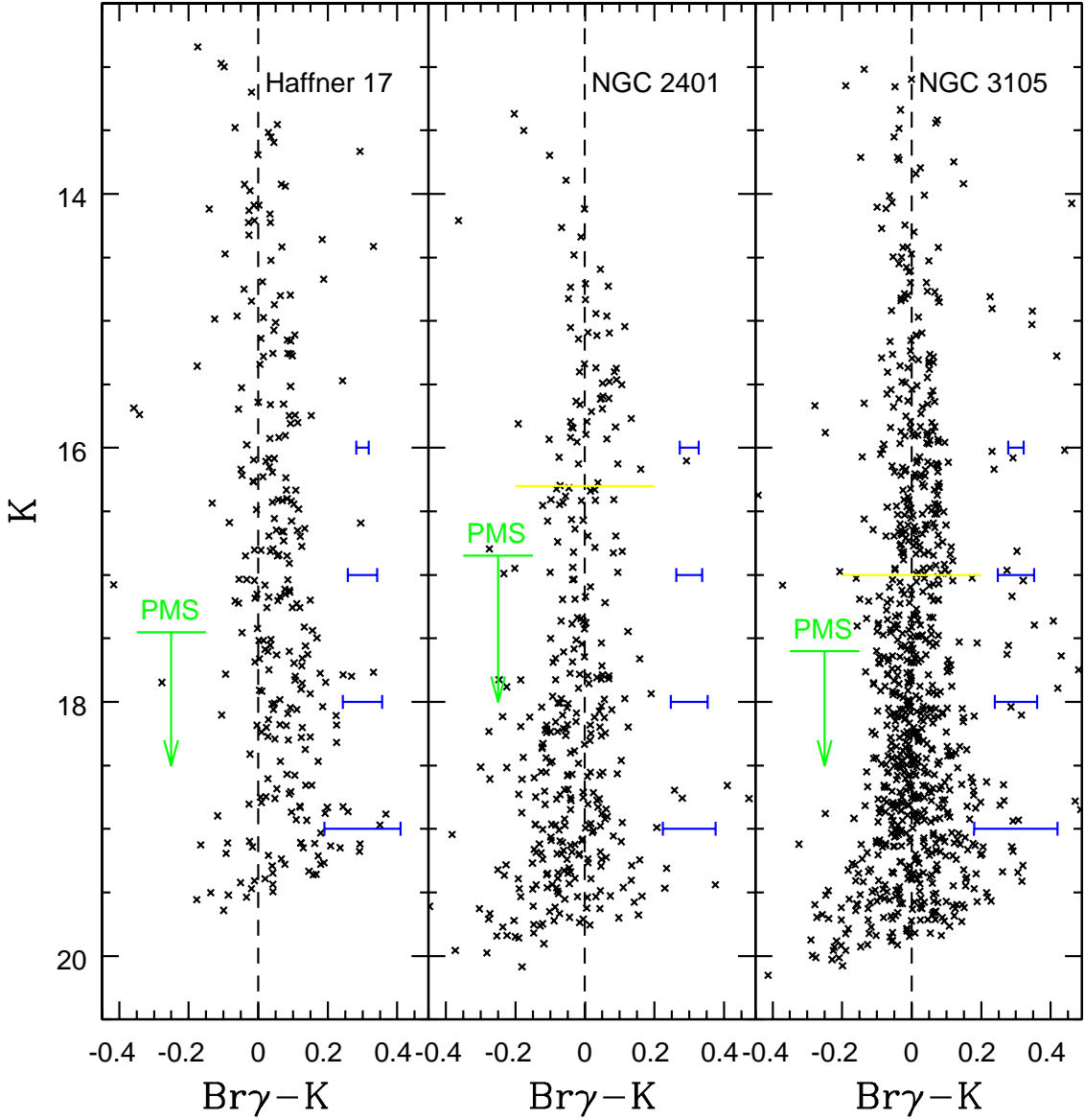}
\caption{$(K, Br\gamma - K)$ CMDs. The dotted line indicates Br$\gamma - K = 0$, 
and the magnitude range where PMS stars are expected is shown for each cluster. 
The blue error bars show the $\pm 1\sigma$ random dispersion in the photometry 
calculated from artificial star experiments. The $(K, 
Br\gamma - K)$ CMD of Haffner 17 does not show obvious changes near the MSCO. 
In contrast, breaks in the Br$\gamma - K$ colors of NGC 2401 and NGC 3105 occur near 
$K = 16.3$ (NGC 2401) and $K = 17$ (NGC 3105), and these breaks are 
indicated with the yellow lines. The artificial star experiments indicate 
that the data for both clusters are complete at these magnitudes, and that the breaks 
are not due to systematic errors in the photometry. Kolmogorov-Smirnov tests indicate 
that the Br$\gamma - K$ distributions of objects in intervals above and below 
these points differ at the 95\%+ confidence level in both clusters.}
\end{figure}

\clearpage

\begin{figure}
\figurenum{7}
\epsscale{0.75}
\plotone{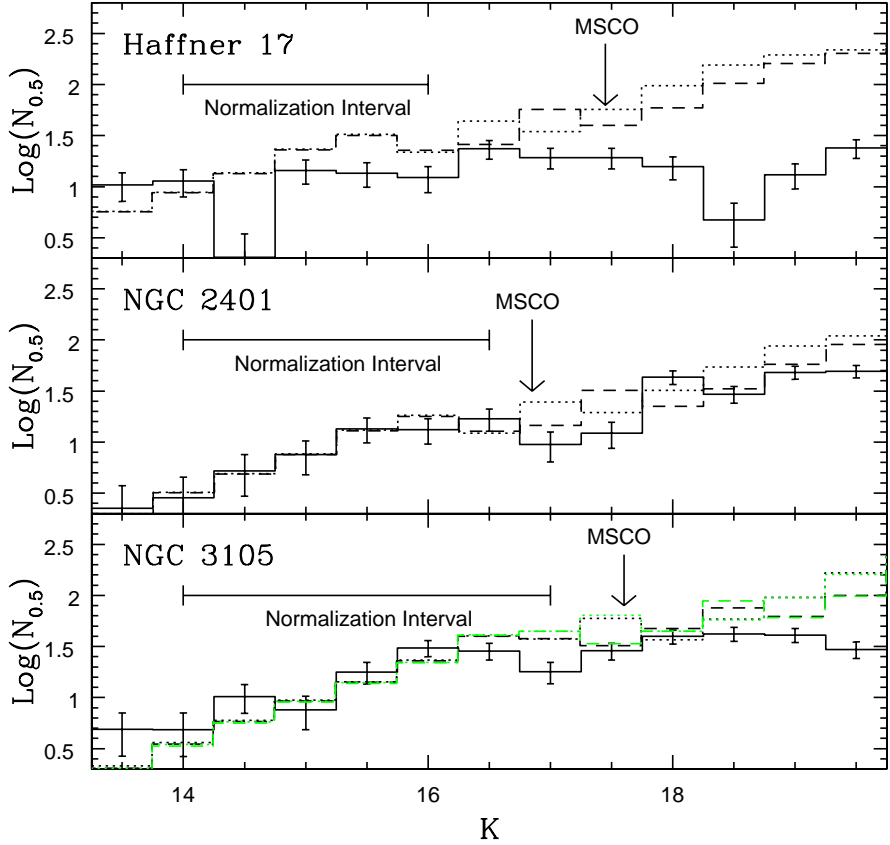}
\caption{$K$ LFs. N$_{0.5}$ is the number of objects per 0.5 magnitude 
interval in $K$, corrected for field stars. {\bf The artificial star 
experiments indicate that the number counts in all three clusters 
are complete for $K < 19.5$} The mean of the upper and lower 
limits of the MSCO is indicated for each cluster. Model LFs with $Z = 0.016$ and ages 
log(t) = 7.4 (dotted line) and 7.6 (dashed line) are also shown. The distance moduli 
and reddenings obtained in Section 4 are used for these comparisons, and the models 
have been scaled to match the observed LFs in the magnitude intervals indicated. 
There is reasonable agreement between the models and the observations 
of NGC 2401 over most magnitudes. However, this is not the case for 
the other clusters. The models do not match the Haffner 17 LF when $K \leq 15$ and 
$K \geq 17$. As for NGC 3105, the models reproduce the slope of the LF when $K 
\leq 16$, but tend to fall above the observed LF at fainter magnitudes. The 
green lines in the bottom panel show the result of adopting a distance modulus 
for NGC 3105 that is 0.15 dex closer than that found 
from the Z = 0.016 sequences, and there is little difference when 
compared with the models that use the larger distance modulus.}
\end{figure} 

\clearpage

\begin{figure}
\figurenum{8}
\epsscale{0.90}
\plotone{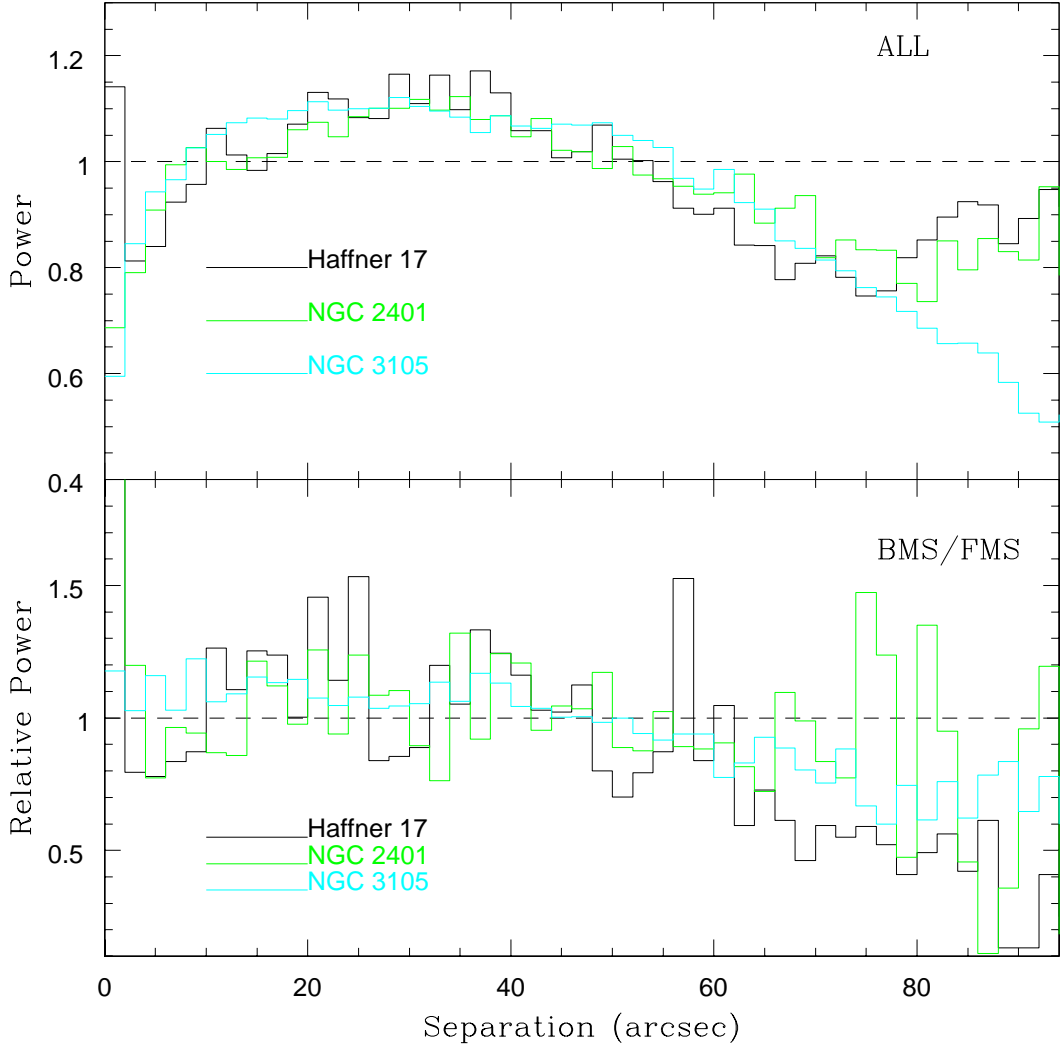}
\caption{TPCFs. The bin-to-bin jitter at intermediate separations provides a rough 
estimate of the statistical noise in these curves (e.g. Davidge 2012). The TPCFs 
of the ALL samples are similar for separations $< 80$ arcsec, except in 
the bin that samples the smallest separation. The peak in the Haffner 17 
TPCF at the smallest separations suggests that 
this cluster may have a larger population of wide binaries than NGC 2401 
and NGC 3105. The ratio of the BMS and FMS SFs, shown 
in the bottom panel, is close to unity for most 
separations in NGC 2401, indicating that the BMS and FMS samples have 
similar clustering properties. However, the ratio of the SFs of Haffner 17 and 
NGC 3105 dips below unity at separations $> 60$ arcsec, indicating that the 
objects in the BMS sample are more centrally concentrated than those in the FMS 
sample. In Figure 9 it is demonstrated that field star contamination likely affects 
the TPCF of both clusters at large separations, although not enough to erase the 
differences in the BMS and FMS distributions.}
\end{figure}

\clearpage

\begin{figure}
\figurenum{9}
\epsscale{0.75}
\plotone{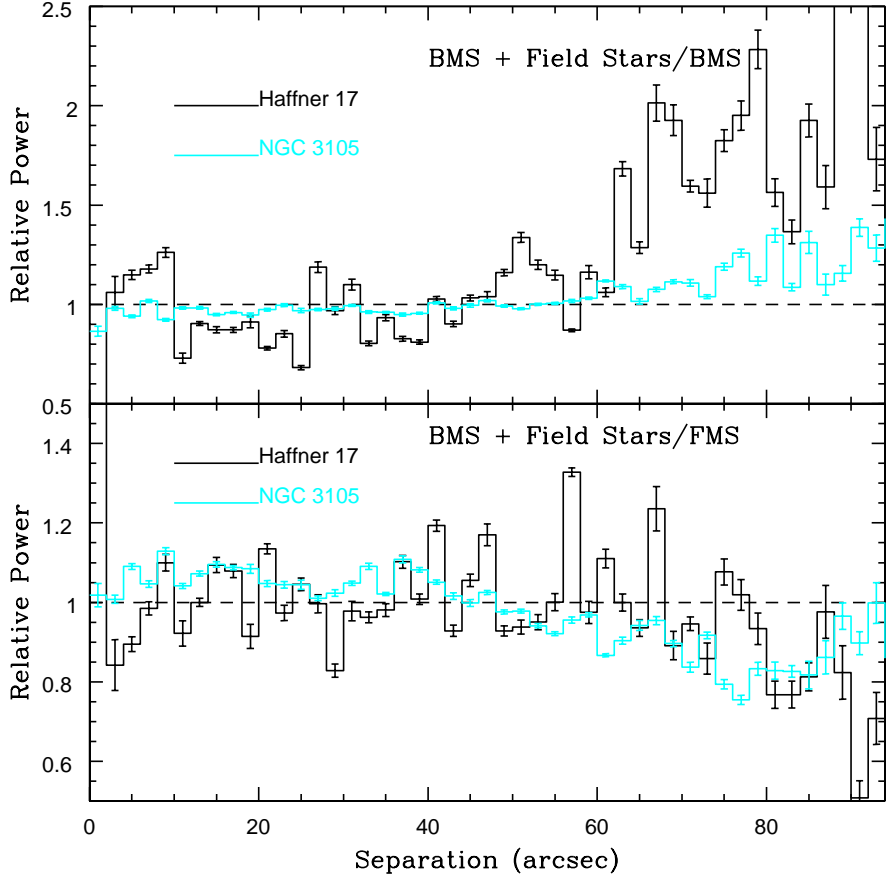}
\caption{Field star contamination and clustering. 
The top panel shows the ratio of the SFs of the BMS $+$ artificial 
field star samples and the measured BMS SFs. The 
BMS$+$field star SF is the average of a number of realizations, 
and the error bars show the dispersion about the mean at each separation. 
Adding a population of uniformly distributed objects has a 
marked impact on the BMS TPCFs of both clusters at separations $> 60$ arcsec. 
The bottom panel shows the ratio of the BMS $+$ field star sample and FMS TPCFs. 
Balancing differences in field star contamination causes the TPCF ratio 
for Haffner 17 to become flatter than in Figure 8, although at 
separations $> 80$ arcsec the TPCF still falls consistently below unity. The objects in 
the FMS sample of NGC 3105 are also more uniformly distributed at large separations 
after adding field stars than those in the BMS sample. The 
members of the BMS and FMS samples that belong to Haffner 17 and NGC 3105 
appear to have different clustering properties, in the sense that the fainter 
members are more uniformly distributed than the brighter, more massive members.}
\end{figure}

\end{document}